\documentclass[A4,12pt]{article} 

\usepackage{graphicx}
\usepackage{dcolumn}
\usepackage{bm}
\usepackage{multirow} 
\usepackage{epstopdf} 
\usepackage{amsfonts}
\usepackage{amsmath}
\usepackage{setspace}
\usepackage{hyperref}
\usepackage[round,authoryear]{natbib}
\hypersetup{colorlinks=true,linkcolor=blue,citecolor=blue,breaklinks={true}}
 
\newcommand{\bc}{\mbox{\boldmath $c$}}
\newcommand{\br}{\mbox{\boldmath $r$}}
\newcommand{\bF}{\mbox{\boldmath $f$}}
\newcommand{\bx}{\mbox{\boldmath $x$}}

\newcommand{\T}{\scriptscriptstyle\rm T }
\font\bbfnt=msbm10

\def\bbR{\mbox{\bbfnt R}}

\newcommand{\mb}[1]{\mbox{\bfseries \itshape #1}}

\begin{document}

\begin{center}
\begin{LARGE}
\bf
Control and Observability Aspects of Phase Synchronization \vspace{1.5cm} 
\end{LARGE}

{\sc  Luis A. Aguirre}
 \vspace{0.25cm}

Departamento de Engenharia Eletr\^onica, Programa de P\'os Gradua\c{c}\~ao em
Engenharia El\'etrica da Universidade Federal de Minas Gerais, Av. Ant\^onio Carlos, 6627,
31.270-901 Belo Horizonte, MG, Brazil. {\tt aguirre@ufmg.br}
 \vspace{0.5cm}

{\sc Leandro Freitas}
 \vspace{0.25cm}

Instituto Federal de Educa\c{c}\~ao, Ci\^encia e Tecnologia de Minas Gerais,
Campus Betim, 32.677-564 Betim MG;
Programa de P\'os-Gradua\c{c}\~ao em Engenharia El\'etrica da
Universidade Federal de Minas Gerais ---  
Av. Ant\^onio Carlos 6627, 31.270-901 Belo Horizonte MG, Brazil.
{\tt leandrofabreu@yahoo.com.br }

\end{center}

\date{\today}

\begin{abstract}
This paper addresses important control and observability aspects of the phase synchronization
of two oscillators. To this aim a feedback control framework is proposed based on which issues related to master-slave
synchronization are analyzed. Comparing results using Cartesian and cylindrical coordinates 
in the context of the proposed framework it is argued that: i)~observability does not play a significant 
role in phase synchronization, although it is granted that it might be relevant for complete synchronization;
and ii)~a practical difficulty is faced when phase synchronization is aimed at but the control action is
not a direct function of the phase error. A procedure for overcoming such a problem is proposed. The
only assumption made is that the phase can be estimated using the arctangent function.
The main aspects of the paper are illustrated using the Poincar\'e
equations, van der Pol and R\"ossler oscillators in dynamical regimes for which the phase is well defined.
\end{abstract}

\section{Introduction}
\label{intro}

The field of synchronization of nonlinear oscillators has witnessed a sharp increase in interest
and number of papers in the last three decades or so. By now, a number of types of synchronization
and ways to achieve them have been reported \citep{boc_eal/02,osi_eal/07}.

It is possibly safe to say that in the earliest days of the synchronization of chaotic oscillators
the emphasis was on complete synchronization (CS) of identical oscillators and the standard way of achieving it was via
multivariate dissipative coupling \citep{fuj_yam/83}. Roughly a decade later, it was shown that
under some conditions CS could be achieved by dissipatively coupling the oscillators via a single variable 
\citep{kap/94}. Another major step in the field was the report of phase synchronization
(PS) as a weaker form of synchronization than CS \citep{ros_eal/96a}. It was shown that PS
very often happened by means of {\it weak}\, dissipative coupling of a single variable in nonidentical oscillators.
Hence, in many
papers, especially the earlier ones, PS was seen simply as an early stage of synchronization, almost a 
``route to CS''. 

Not before long, it became clear that PS was a relevant type of synchronization {\it per se}. 
In fact, in many engineering applications it is PS that matters \citep{piq/11}, however 
the ``traditional'' dissipative coupling is still commonly used. It has been pointed out 
in \citep{bel_eal/05} that such a way of coupling might not be effective in PS applications.

An interesting and related problem in the context of synchronization attained by
monovariable coupling is that of deciding which variable to use. To the best of the authors'
knowledge there is no general choice. Nonetheless, it was shown that for some investigated systems
the variables used in bidirectional dissipative coupling that achieved synchronization (CS and PS)
for lower values of coupling gain often coincided with those variables that provided greater
observability of the dynamics, although the results are also influenced by 
the dynamical regime \citep{let_agu/10}. A similar approach was recently extended to
investigate how controllability might influence the establishment of CS and PS to an 
external signal, however no general relationship was found \citep{agu_let/16}. In a similar
vein, possible relations between classes of synchronization and observability have been
investigated in \citep{sen_eal/16}, where the authors conclude that although some
relation exists, observability only explains synchronization partially.

The relationship between observability and the ability to synchronize two oscillators by
monovariable dissipative coupling can be understood in the following terms. For the sake
of argument, let us assume that $\bx =[x,\, y,\, z]^{\T}$ and that the coupling is done using the $x$ variable, hence, 
for the slave oscillator a (linear) dissipative coupling term of the form $k_1[x_2(t)-x_1(t)]$, where
the subscripts indicate the different oscillators (1 for the slave oscillator and 2 for the master), will
be added to the first equation of the slave oscillator \footnote{In case of bidirectional coupling, the term 
$k_2[x_1(t)-x_2(t)]$ will be added to the first equation of the second oscillator, and the oscillators are no
longer called master or slave.}. Now assume 
that CS is achieved, then $|x_2(t)-x_1(t)|\rightarrow 0$, $|y_2(t)-y_1(t)|\rightarrow 0$ and
$|z_2(t)-z_1(t)|\rightarrow 0$, although {\it the only}\, signal used from the master oscillator was $x_2(t)$.
Hence, somehow the knowledge of $x_2(t)$ is sufficient to figure out the full state of the
second oscillator. Hence the first oscillator plays the role of an observer of the second one. 
This is typically the case when $x$ conveys good observability of the
dynamics. Conversely, if $x$ is poor from an observability point of view, to synchronize
the oscillators by coupling using such a variable is more difficult \citep{let_agu/10}.
Hence there seems to be an acknowledged connection between observability and 
synchronization at least in some contexts \citep{par_eal/96,may_amr/99,fre_eal/05,sed_eal/12}.

How does all this translate to the case of PS? As for the coupling, it has already been
pointed out that dissipative coupling is not always effective to achieve PS \citep{bel_eal/05},
but why? Also, could it be that a more fundamental issue in synchronization problems -- 
at least in the case of PS -- is not so much the level of (state) observability
provided by the variable used in the coupling but rather the amount of information concerning
the phase conveyed by such a variable? In order to address such questions PS is set in
the context of a control problem. This framework -- the lack of which has been pointed out in 
\citep{bel_eal/05} -- is used to investigate PS and to point out the need for new tools.
Using this framework it will be argued that a practical problem is that of a mismatch that
arises between the controlled and manipulated variables. A procedure to overcome this
shortcoming is proposed.

This paper is organized as follows. In Section~\ref{fcf} the proposed feedback control framework
is presented. The main points and definitions of the paper are provided in that section. In particular,
the {\it control action and manipulated variable mismatch}\, problem is pointed out in Sec.\,\ref{secrc},
and a solution to that problem is proposed in Sec.\,\ref{sol}.
The numerical results using benchmark oscillators in two systems of coordinates -- Cartesian and cylindrical --
are presented in Section~\ref{nr}. The results are discussed in Section~\ref{dc} where the main conclusions
and suggestions for future work are also provided.

\section{Feedback Control Framework}
\label{fcf}

\begin{figure}[!ht]
		\centering
		\includegraphics[width=0.8\textwidth]{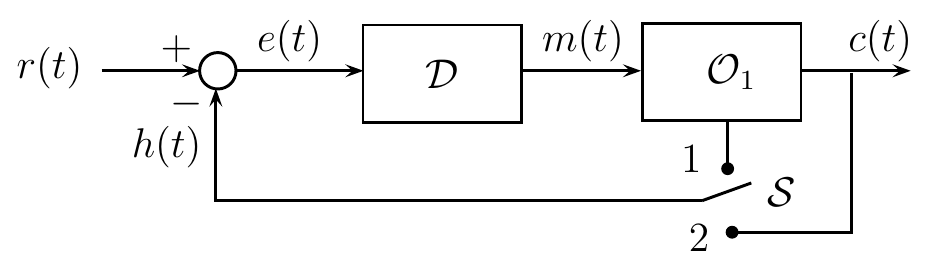}
		\caption{\label{contloop}Typical control loop.}
\end{figure}

In order to motivate the main point of this paper we start by describing the
typical control loop shown in Figure~\ref{contloop}. The input $r(t)$ is the reference or
set point. The feedback signal $h(t)$ is compared to the reference and an ``error'' signal 
$e(t)$ is produced based on which the controller ${\cal D}$ yields the \emph{control action} (CA) $m(t)$ 
to drive the plant ${\cal O}_1$, by changing the corresponding \emph{manipulated variable} (MV) within ${\cal O}_1$.
The signal at the righthand of the loop, $c(t)$, is the \emph{controlled variable}.

In most control problems, the controlled variable $c(t)$ is fedback and compared to the
reference. In Figure~\ref{contloop} this can be attained by placing the switch ${\cal S}$ in
 position 2, and hence
$h(t)=c(t)$. In such a situation, the controller ${\cal D}$ is supposed to drive the plant until
$e(t)=r(t)-c(t)$ remains within acceptable bounds. In the sequel the basic control scheme just
described will be used to analyze two different types of synchronization.

\subsection{Complete synchronization}

Consider a generic 2D periodic oscillator with control inputs $m_x(t)$ and $m_y(t)$
\begin{equation}
  \label{gen}
{\cal O}_1:  \left\{
    \begin{array}{lll}
      \dot{x}_1 & = f_x(x_1,\,y_1) + m_x(t)\\
      \dot{y}_1 & =  f_y(x_1,\,y_1) + m_y(t),\\
    \end{array}
  \right.
\end{equation}
Assume that we choose $r_x(t)=x_2(t)$ 
and $m_y=0$, where $x_2(t)$ is the first state variable from another oscillator (the master). 
Suppose we want $x_1(t)$ to follow the reference, then we choose $c(t)=x_1(t)$,
place ${\cal S}$ in position 2, and use, say, dissipative (or diffusive) coupling $m_x(t)=k_x[r_x(t)-x_1(t)]$.
Often there is a range of values of $k_x$ (and $m_y=0$) for which $|e(t)|\rightarrow 0$ is achieved. In 
principle, coupling could also be achieved using the $y$ variable, that is taking $k_x=0$, $r_y(t)=y_2(t)$, $c(t)=y_1(t)$ 
and $m_y(t)=k_y[r_y(t)-y_1(t)]$ it is often possible to get $|e(t)|\rightarrow 0$ over a range of values of $k_y$.

The scheme just described has been pursued since the early days of synchronization of chaotic -- 3D --
oscillators \citep{fuj_yam/83}, where coupling was dissipative, as described in the preceding paragraph, although
controlling all state variables, that is, the control aim was to make the full
state vector $\bc(t)=\bx_1(t)$ follow the reference $\br(t)=\bx_2(t)$. An early paper in which
only one state variable was fedback (like shown in Figure~\ref{contloop}) is \citep{kap/94} where 
complete synchronization was also the goal.

{\bf Remark 1.} An important aspect to notice is that in (\ref{gen}) the \emph{control action} $m_s(t)$ is added
to the corresponding equation, that is, to the equation that defines the dynamics of the \emph{manipulated 
variable}, $s\in\{x,\,y\}$. When the \emph{control action} is a function of the error in the \emph{manipulated variable},
as in this case, we have a favorable situation from a control point of view.

{\bf Remark 2.} Another common problem is the PS to an external force -- usually periodic -- 
which is simply added
to one of the equations of the oscillator ${\cal O}_1$. Such is an {\it open-loop}\, type of operation
in which the switch ${\cal S}$ remains opened, hence $e(t)=r(t)$. An early example of this was reported in
\citep{sto/92} for the case of the R\"ossler system with $r(t)$ an impulsive force acting on the $y$ 
component at a constant frequency.

\subsection{Phase synchronization}

In 1996, Rosenblum and co-workers reported a different type of synchronization: phase synchronization
\citep{ros_eal/96a}. They showed that the R\"ossler system 
with $m_x(t)=k_x[r_x(t)-x_1(t)]$ and $m_y(t)=0$ would satisfy the condition for phase synchronization,
also known as non-strict phase locking \citep{osi_eal/07}:
$|\phi_2(t)-\phi_1(t)|<{\rm const}$, even though $e(t)=r_x(t)-x_1(t)$ did {\it not}\, converge to zero, but
rather $r_x(t)$ and $x_1(t)$ remained practically uncorrelated. In the aforementioned paper the authors
reported on two cases: i)~$r_x(t)$
was the $x$ variable of another R\"ossler oscillator, and ii)~$r_x(t)$ was the $u$ 
variable of a bidirectionally coupled Mackey-Glass differential-delay system. 

In what follows we discuss different choices for the reference signal and therefore for the
control action, and this related to the mathematical representation of oscillator ${\cal O}_1$. 
We start with the most commonly found situation.

\subsubsection{Cartesian coordinates}
\label{secrc2}

This is the most common situation found in the literature and it is not difficult to see why. As pointed out in
Sec.\,\ref{intro}, this situation emerges from CS problems where the aim is to attain $\bx_1=\bx_2$. Since the
PS was reported, many works have focused on such a phenomenon, though retaining the original and more general setting.
So here we analyze PS in this ``CS-oriented'' setting, that is, $r_s(t)=s(t),~s\in\{x,\,y$\}. In what follows we 
choose $r_x(t)=x_2(t)$, although $r_y(t)=y_2(t)$ could be used instead.
 
In PS investigations, it would be natural to chose the phase as the controlled variable, that is, $c(t)=\phi_1(t)$ although the
reference signal is a state variable $r_x(t)=x_2(t)$ and not a phase signal, because of the (inherited) general CS setting. 
In order to produce an error signal
that makes sense,\footnote{Ideally this signal should be a function of the phase error, e.g. $e(t)=\phi_2(t)-\phi_1(t)$.}   
${\cal S}$ is connected to 1 and the first state variable of ${\cal O}_1$ is fedback, hence
$h(t)=x_1(t)$, rendering $e(t)=x_2(t)-x_1(t)$. One of the simplest controllers is proportional to the error,  
and therefore the control action is $m_x(t)=k_x[x_2(t)-x_1(t)]$, although it might not be the most
effective choice \citep{bel_eal/05}. According to the nomenclature defined $m_x(t)$ is added to the first equation of ${\cal O}_1$
(as in Eq.\,\ref{gen}) and thus  the welcome matching between the control action and the manipulated variable is attained. 
However, the resulting control problem in this case is somewhat unconventional, with ${\cal S}$ connected to 1
and with the \emph{reference} and \emph{error} of a different type than the \emph{controlled variable}.

{\bf Remark 3.} The condition for PS is that $|\phi_2(t)-\phi_1(t)| < {\rm const}$. However the control actuates in a
way to reduce the error $e(t)=x_2(t)-x_1(t)$. Therefore the key to the success of pursuing PS using a CS-oriented 
setting is that $x_2(t)$ and $x_1(t)$ should be good {\it proxy variables}\, for $\phi_2(t)$ and $\phi_1(t)$, respectively.

\subsubsection{Cylindrical coordinates}
\label{seccc}

In order to focus on the phase dynamics of an oscillator a common procedure is to reduce the oscillator
to a {\it phase oscillator}, which has an explicit equation for the phase, as in the famous model
proposed by Kuramoto \citep{kur/75,dor_bul/14}. \citet{jos_mar/01} showed that it is possible to reduce
a chaotic oscillator to a phase oscillator by means of an appropriate change of coordinates, which exists
for most phase-coherent systems. Hence the use of phase oscillators solves two
important problems. First, because the phase is readily available, there is no need to use proxy
variables for it. Secondly, the \emph{control action} can be added directly to the equation that describes
the phase dynamics thus avoiding any mismatch between the \emph{control action} and the \emph{manipulated variable}.
This will be detailed below.

Now let us assume that the aim is to achieve PS of two oscillators. For the sake of argument we only address
the case of unidirectional coupling and refer the reader to Figure~\ref{contloop}. From an {\it ideal control}\, point of view
in this problem the reference should be a phase signal, hence $r_\phi(t)=\phi_2(t)$, the controlled variable should be the corresponding
phase of ${\cal O}_1$, hence $c(t)=\phi_1$, switch ${\cal S}$ should be connected to terminal 2, and 
$e(t)$ then becomes the {\it phase error}. In this framework the control goal is that $|e(t)|=|\phi_2(t)-\phi_1(t)|<{\rm const}$
for all $t>t_0$. In the case of dissipative (or diffusive) coupling, the controller can be a pure gain $k_\phi$, such that
the control action is $m_\phi(t)=k_\phi[r_\phi(t)-\phi_1(t)]$, where $r_\phi(t)=\phi_2(t)$.\footnote{A common nonlinear
and $2\pi$-periodic control action is $m_\phi(t)=k_\phi \sin [r_\phi(t)-\phi_1(t)]$. In this paper we use
a $2\pi$-periodic control action $m_\phi(t)=k_\phi {\tt wraptopi} [r_\phi(t)-\phi_1(t)]$, where {\tt wraptopi}
wraps angles to the range $(-\pi,\,\pi]$.} 

In order to clearly see this ideal framework, let us rewrite system (\ref{gen})  in
cylindrical\footnote{In 2D this boils down to polar coordinates. Because in this paper 2D and 3D oscillators are
considered, the term {\it cylindrical} coordinates will be used.} coordinates:
\begin{equation}
\label{cyl}
{\cal O}_1:  \left\{
    \begin{array}{lll}
      \dot{\rho}_1 & = g_\rho (\rho_1,\, \phi_1 ) + m_\rho(t)  \\
      \dot{\phi}_1 & =g_\phi (\rho_1,\, \phi_1 ) + m_\phi(t)  .
    \end{array}
  \right.
\end{equation}

\noindent
The advantage of representing
${\cal O}_1$ as in (\ref{cyl}) is twofold: first, numerical integration directly provides the phase, which in this
problem is the controlled variable $c(t)$; second, it permits actuating directly on the equation that describes
the phase dynamics. 
If PS is intended we choose $c(t)=\phi_1(t)$, $r_\phi(t)=\phi_2(t)$ 
and $m_\rho(t)=0$, where $\phi_2(t)$ is the phase from the master oscillator.
Connecting ${\cal S}$ to position 2, and using dissipative coupling as $m_\phi(t)=k_\phi[r_\phi(t)-\phi_1(t)]$
this scheme is quite successful in attaining PS, as will be shown.

\subsubsection{Control action and manipulated variable mismatch}
\label{secrc}

For the sake of completion, we investigate a rather awkward situation: ${\cal O}_1$ will be represented in
Cartesian coordinates but the use of proxy variables is avoided by choosing the reference $r_\phi(t)=\phi_2(t)$,
and hence the control action will be $m(t)=k_\phi[\phi_2(t)-\phi_1(t)]$.
A second difficulty is that $m(t)$ is connected to the phase error, but because there is no equation for the
phase dynamics, regardless to which equation $m(t)$ is added, there will be a 
{\it control action and manipulated variable (CA-MV) mismatch}. The one positive aspect of this
strange configuration is that no proxy variables are employed.

In a sense this control problem is a {\it transition}\, between the ones described in
sections \ref{secrc2} and \ref{seccc}. Here we still aim at PS and
to emphasize this we still assume that $c(t)=\phi_1(t)$ and $r_\phi(t)=\phi_2(t)$
and we do {\it not}\, resort to the helpful cylindrical coordinates (\ref{cyl}) but rather consider the system
to be in Cartesian coordinates as in (\ref{gen}). Since $\phi_1$ is not a state variable in this
representation, we do not explicitly feedback that signal, instead we place the switch ${\cal S}$ in 
position 1 and feedback a function of the states of the system. It is assumed that
$h(t)=\tan^{-1}(y_1/x_1)$ is a good estimate for the phase\footnote{The arctangent function is considered
with two arguments: {\tt atan2($y,\,x$)}, where the angle information can be obtained in the four
quadrants.} and hence the error $e(t)$ is the phase difference. This is almost
equivalent to the case investigated in Sec.\,\ref{seccc}, if it was not for the fact that here we
face a CA-MV mismatch.

\subsection{Other synchronization schemes}

Here we briefly mention that the proposed framework can be used to understand other synchronization schemes
found in the literature. In what follows two PS are considered as illustrative examples.

As a first example, we quote the phase-locked loop (known by the acronym PLL) 
which first appeared early in the 20th century \citep{hsi_hun/96}. The main elements of a PLL are a
phase detector, a low-pass filter and a voltage-controlled oscillator \citep{piq/11}. This scheme has
two of the principal benefits argued so far, namely, given two sinusoidal oscillations, the phase error
is detected {\it and}\, the oscillator frequency is changed (controlled) proportionally to such an error,
hence avoiding the CA-MV mistmatch. 

Another interesting scheme to guarantee PS and that closely resembles a PLL
has been proposed in \citep{bel_eal/05} where the controller output is a function of the {\it phase difference}, i.e. $m(t)=F[e(t)], ~e(t)=\phi_2(t)-\phi_1(t)$, that
is {\it added to frequency-related parameters}, which is a way of avoiding the CA-MV mismatch although it might be
difficult to implement in practice.

In summary, the success of the two PS schemes mentioned in this section stems mainly from two facts related
to the control action i)~it is a function of the phase difference, and ii)~it directly influences the instantaneous
frequency of ${\cal O}_1$, avoiding CA-MV mismatch.

\subsection{Observability requirements}

The observability of linear dynamical systems was introduced by Rudolf Kalman \citep{kal/60} and
has been extended and investigated for nonlinear system in many works, as for instance \citep{her_kre/77}.
This concept has proven relevant in a number of problems in nonlinear dynamics \citep{par_eal/14} where observability
coefficients $\delta_s$ have been proposed as average condition numbers of observability matrices 
${\cal O}_s$ for the system when only variable $s(t)=h(\bx)$ is recorded from the system \citep{let_eal/05pre}.

The concept of observability can be understood as follows. Assume that the dynamical system under study is 
$\dot{\bx} (t) = \bF ( \bx (t) )$, where $t$ is the time, $\bx \in \bbR^n$ the state vector and $\bF$ 
the nonlinear vector field. Let  $s(t)$ be a scalar time series obtained using 
a measurement function $h: \bbR^n \rightarrow \bbR$, that is $s(t) = h(\mb{x}(t))$. The system
is said to be {\it observable}\, if, given $s(t)$, it is possible to find the initial condition $\bx_0$ for
finite $t$. In more practical terms, if the pair ($\bF,\,s$) is observable, then it is possible to determine the 
state $\bx(t)$ from the recorded time series $s(t)$.

Now it becomes clear that because the goal in CS is to attain $\bx_1=\bx_2$ and if only one state variable is used,
say $x$, then it makes sense to require that ($\bF_2,\,x_2$) be observable, because from the reference
signal $r_x(t)=x_2(t)$ the controlled oscillator ${\cal O}_1$ should be able to ``see'' the full state
of the master oscillator $\bF_2$ in order to follow it. No wonder there is significant correlation between
observability and CS via dissipative bidirectional coupling \citep{let_agu/10} or by complete
substitution \citep{pec_car/90}, as pointed out in \citep{agu_let/16}.

On the other hand, if PS is aimed at and if the framework in Sec\,\ref{seccc} is considered, {\it there is no need to require that
($\bF_2,\,\phi_2$) be observable}\, because there is no need to ``see'' (reconstruct) the entire state vector
from the recorded $\phi_2(t)$. Hence the requirements on observability in synchronization problems
will have some relevance if the aim is CS or possibly lag synchronization (LS) \citep{ros_eal/97}, but 
might not be the right feature to ask for if the goal is PS. This ``decoupling'' between PS and observability
will be illustrated with simulations in Section~\ref{nr}, after proposing a solution the the  CA-MV mismatch problem.

\section{A Solution to the CA-MV mismatch problem}
\label{sol}

In this section a solution to the CA-MV mismatch problem pointed out in Sec.\,\ref{secrc} is proposed.
Without loss of generality it is supposed that the slave oscillator is
described in 2D Cartesian coordinates with possible control actions $m_x$ and $m_y$ actuating directly on $\dot{x}_1$ and $\dot{y}_1$ equations, respectively. It is assumed that the phase of the slave oscillator can be adequately estimated by $\phi_1=\tan^{-1}(y_1/x_1)$
and that $m_x$ or $m_y$ should be such as to reduce the phase error $e(t)=r_{\phi_2}(t)-\phi_1(t)$, where the reference
signal $r_{\phi_2}(t)=\tan^{-1}(y_2/x_2)$ is the phase of the master oscillator.

The ideal situation occurs when the CA actuates directly on the desired \emph{manipulated variable} (MV) dynamics, that is $\dot{\phi}_1$.
This is {\it not}\, the case here as the slave oscillator is assumed to be in Cartesian coordinates. The rationale behind the solution
to be developed is to {\it find suitable control actions $m_x$ and $m_y$ in Cartesian coordinates that would effectively influence
the phase dynamics}. To achieve this, the phase dynamics of the master oscillator is written thus
\begin{eqnarray}
  \label{eq:dphidx}
  \dot{\phi_1} &=& \frac{\partial \phi_1}{\partial x_1} \dot{x}_1 + \frac{\partial \phi_1}{\partial y_1} \dot{y}_1 \nonumber\\
  &=& \frac{-y_1}{x_1^2+y_1^2} (\dot{x} + m_x) + \frac{x_1}{x_1^2+y_1^2} (\dot{y} + m_y) \nonumber\\
  &=& \left[ \frac{-y_1}{x_1^2+y_1^2} \dot{x} + \frac{x_1}{x_1^2+y_1^2} \dot{y}\right] + \frac{-y_1 m_x + x_1 m_y}{x_1^2+y_1^2} \nonumber\\
  &=& \nu(\boldsymbol{x}) + \frac{-y_1}{x_1^2+y_1^2} m_x + \frac{x_1}{x_1^2+y_1^2} m_y,
\end{eqnarray}
where the variables $(\dot{x},\,\dot{y})=(f_x(x_1,\,y_1),\,f_y(x_1,\,y_1))$ are the time derivatives of the uncoupled oscillator 
(see Eq.\,\ref{gen}) and 
the term in brackets -- called $\nu(\boldsymbol{x})$ -- is its natural frequency.
In the second line the following result was used
\begin{eqnarray}
  \frac{\partial~\tan^{-1} (y/x)}{\partial \bx} = \left[  \frac{-y}{x^2+y^2} ~ \frac{x}{x^2+y^2} \right]^{\T} . \nonumber
\end{eqnarray}
%


To see the consequences of the CA-MV mismatch, two situations will be considered in turn.

\begin{itemize}
\item[(a)]
$m_x=k_\phi[r_{\phi_2}(t)-\phi_1(t)]$ and $m_y=0$. From  (\ref{eq:dphidx}) the phase dynamics can
be described thus:
\begin{equation}
  \label{eq:dphidx1}
  \dot{\phi_1} = \nu(\boldsymbol{x}) + \left(\frac{-y_1 }{x_1^2+y_1^2} \right) k_\phi\left[r_{\phi_2}(t)-\phi_1(t)\right].
\end{equation}
Assuming $k_\phi>0$, whenever the term in parenthesis is negative there will be a positive feedback loop, 
reinforcing the phase error. 

\item[(b)] $m_y=k_\phi[r_{\phi_2}(t)-\phi_1(t)]$ and $m_x=0$. As before, from~(\ref{eq:dphidx}) the
phase dynamics are:
\begin{equation}
  \label{eq:dphidx2}
  \dot{\phi_1} = \nu(\boldsymbol{x}) + \left(\frac{x_1 }{x_1^2+y_1^2} \right) k_\phi\left[r_{\phi_2}(t)-\phi_1(t)\right],
\end{equation}
and feedback will be positive whenever the term in parenthesis is negative.
\end{itemize}

Hence, as the trajectory travels through state space, there will be regions of positive feedback where such
a simple scheme will fail to keep phase error small.
A solution to this scenario, which is a consequence of the CA-MV mismatch, is to choose 
\begin{eqnarray}
\label{choice}
m_x &=& -y_1 k_{\phi_x}[r_{\phi_2}(t)-\phi_1(t)] \nonumber \\
m_y & =& x_1 k_{\phi_y}[r_{\phi_2}(t)-\phi_1(t)].
\end{eqnarray}

\noindent
Hence, the phase dynamics will be of the form
\begin{equation}
  \label{eq:dphidx3}
  \dot{\phi_1}  =  \nu(\boldsymbol{x}) + \tilde{k}_\phi(t)[r_{\phi_2}(t)-\phi_1(t)],~\tilde{k}_\phi(t)>0,
\end{equation}
which always results in negative feedback. In (\ref{eq:dphidx3}) $\tilde{k}_\phi(t)$ includes the state variables $x_1(t)$, $y_1(t)$ and
the constant gains $k_{\phi_x}$ and $k_{\phi_y}$. If $k_{\phi_x}=k_{\phi_y}=k_\phi$, then $\tilde{k}_\phi(t)=k_\phi$.

The only assumption that leads to (\ref{choice}) is that the phase can be described by 
arctangent function of a quotient of two state variables. The procedure can be applied to 3D oscillators as
long as the phase can be computed taking the arctangent of the quotient of two of the state variables.
Hence such a procedure is applicable to a large class of systems. 
Results for the R\"ossler oscillator will be presented in Sec.\,\ref{rc2}.

\section{Numerical Results}
\label{nr}

In this section we numerically investigate the PS of oscillator ${\cal O}_1$ with the
reference signal $r(t)$ provided by a master oscillator in order to support the analysis and claims
put forward in the previous sections.
The oscillators considered are: Poincar\'e, Van der Pol and  R\"ossler, for which representation in
Cartesian and cylindrical coordinates are readily available in the literature and also because
they display phase coherent dynamics. This is important to test the proposed framework because
other relevant practical problems, such as defining and estimating the phase for phase noncoherent
dynamics, are avoided.

In what follows potential control actions (coupling terms) will be indicated in the system equations.
In transforming the equations to cylindrical coordinates, only the autonomous part of the
original equations were considered and the potential control action was added later. In all
simulations the transients were removed from the analysis.

The equations for the Poincar\'e system are \footnote{As for the origin of this system 
the following quotation is relevant ``Since the first description of two-dimensional radially 
symmetric differential equations with limit cycles was given by Poincar\'e, we propose that these 
systems be called {\it Poincar\'e oscillators}'' \cite[pp.\,25]{gla_mac/88}. However, It is pointed out
that the denomination is used for a class of two-dimensional differential equations, and the specific
oscillator in (\ref{poir}) is not considered in that reference.}\,\citep{bel_eal/05}:
\begin{equation}
  \label{poir}
 \left\{
    \begin{array}{lll}
      \dot{x}_1 & =-\omega y_1 -\lambda(x_1^2+y_1^2-p^2)x_1 + m_x(t)\\
      \dot{y}_1 & = \omega x_1 -\lambda(x_1^2+y_1^2-p^2)y_1  +m_y(t) ,\\
    \end{array}
  \right.
\end{equation}

\noindent
with $(\omega ,\, p,\, \lambda)=(1,\,1,\,0.5)$ and which can be represented in cylindrical coordinates as:
\begin{equation}
  \label{poip}
 \left\{
    \begin{array}{lll}
     \dot{\rho}_1 & =\lambda \rho_1 (p^2- \rho_1^2)  \\
      \dot{\phi}_1 & = \omega + m_\phi (t).
    \end{array}
  \right.
\end{equation}

The van der Pol oscillator \citep{vdp/27} can be described as:
\begin{equation}
  \label{vdpr}
\left\{
    \begin{array}{lll}
      \dot{x}_1 & =\omega y_1 +  m_x(t)\\
      \dot{y}_1 & = -\omega x_1 -\mu(\beta x_1^2-1)y_1  +m_y(t)  ,\\
    \end{array}
  \right.
\end{equation}

\noindent 
with $(\omega,\, \mu,\, \beta)=(1,\,0.1,\,1)$, in which $\mu$ indicates the nonlinearity of the system. 
Defining $x_1=\rho_1 \cos \phi_1$ and  $y_1=\rho_1 \sin \phi_1$, the oscillator can
be described in cylindrical coordinates as:
\begin{equation}
  \label{vdpp}
 \left\{
    \begin{array}{lll}
     \dot{\rho}_1 & =-\mu (\beta \rho_1^2 \cos^2 \phi_1 -1) \rho_1 \sin^2 \phi_1 \\
      \dot{\phi}_1 & = -\omega -\mu(\beta \rho_1^2 \cos^2 \phi_1 -1)\sin \phi_1 \cos \phi_1+ m_\phi (t) ,
    \end{array}
  \right.
\end{equation}

\noindent 
which is valid $\forall \bx \neq {\bf 0}$, and the phase is assumed to be $\phi_1=\tan^{-1} (y_1/x_1)$.

The R\"ossler system \citep{ros/76} in Cartesian coordinates with control inputs is described as:
\begin{equation}
  \label{roso76}
 \left\{
    \begin{array}{lll}
      \dot{x}_1 & = -\omega y_1 -z_1 + m_x(t) \\
      \dot{y}_1 & = \omega x_1+ay_1 +m_y(t) \\
      \dot{z}_1 & = b+z_1(x_1-c) +m_z(t) , \\
    \end{array}
  \right.
\end{equation}
and $(a,\, b,\, c,\,\omega)=(0.398,\,2.0,\,4.0,\,1.0)$.
For these parameters, considering the control action in the $x$ variable, $m_y(t) = m_z(t) = 0$, with 
$m_x (t) = k_x [ r_x(t) - x_1(t) ]$ and $k_x=1.5$, $|e(t)|\to0$ is achieved.
The R\"ossler system can be rewritten in cylindrical coordinates with control action in the
second equation as \citep{ros_eal/97}:
\begin{equation}
\label{roscyl}
 \left\{
    \begin{array}{lll}
      \dot{\rho}_1 & = a\rho_1\sin^2 \phi_1 -z_1 \cos \phi_1 \\
      \dot{\phi}_1 & = \omega + a\sin \phi_1 \cos \phi_1 + (z_1/\rho_1) \sin \phi_1 + m_\phi (t) ,\\
      \dot{z}_1 & = b+z_1(\rho_1\cos \phi_1 - c) .
    \end{array}
  \right.
\end{equation}

\noindent
For phase coherent dynamics, $\phi_1$ in (\ref{roscyl}) has a clear interpretation of phase, as
it can be computed thus $\phi_1=\tan^{-1} (y_1/x_1)$.

These oscillators will be considered in the light of Sections \ref{secrc2} to \ref{secrc}. To evaluate the quality of
PS the {\it mean phase coherence}\, will be used. This has been defined as \citep{mor_eal/00}:
\begin{eqnarray}
\label{pc}
R=\sqrt{\!\langle \sin \!|\phi_2(t)\!-\!\phi_1(t)| \rangle^2 \!+\! \langle \cos \!|\phi_2(t)\!-\!\phi_1(t)| \rangle^2} ,
\end{eqnarray}

\noindent
where $\langle \cdot \rangle$ is time average. $R=1$ is achieved for strict phase locking, that
is, for $|\phi_2(t)-\phi_1(t)|=$const.

\subsection{Cartesian coordinates}
\label{cc}

In this section we simulate the most commonly found situation: the oscillators are represented in Cartesian
coordinates, coupling is dissipative, and $r_s(t)=s_2(t),~s\in~\{x,\,y,\,z\}$, which means that the control action 
for the first equation is $m_x(t)=k_x[x_2(t)-x_1(t)]$ and the error signal $e(t)=x_2(t)-x_1(t)$ is 
in the {\it state variable}\, $x$, not in the phase. Analogous definitions are used for the other equations.
Therefore there is no CA-MV mismatch. 
The control action is added to one equation at a time. Also, PS will be assumed as the goal.

\begin{figure}
		\centering
		\includegraphics[width=0.8\textwidth]{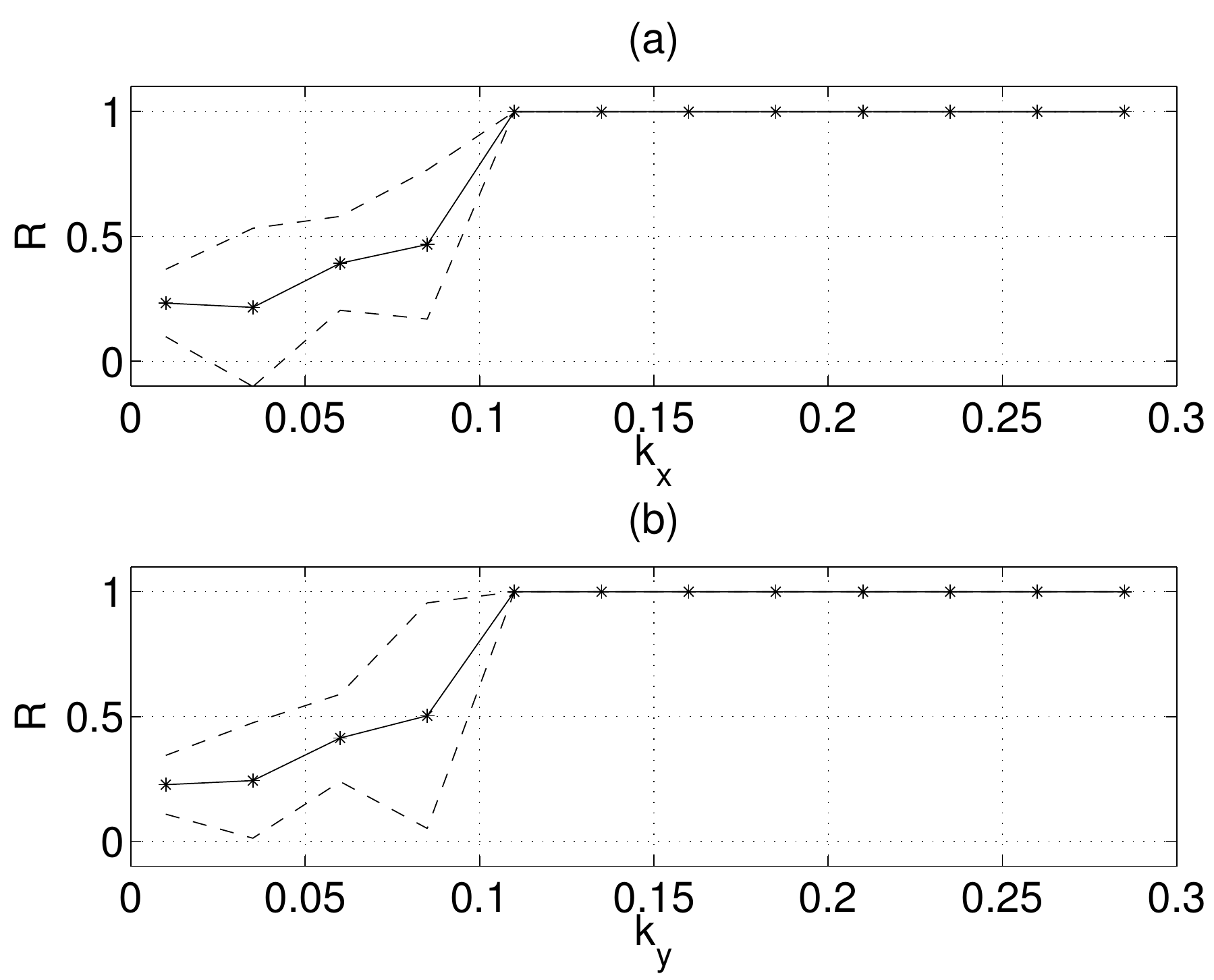}
		\caption{\label{poinc_kxyMCrec}Phase coherence $R$ (\ref{pc}) for the Poincar\'e oscillator in Cartesian 
		coordinates with  (a)~$r_x(t)=x_2(t)$ and  $k_y=0$, (b)~$r_y(t)=y_2(t)$ and $k_x=0$.
		Results for 10 Monte Carlo runs with random initial conditions. The plot shows the mean values with
		asterisks and the $\pm 2\sigma$ band with dashed lines. The parameters for ${\cal O}_1$ are
		$(\omega,\, p,\, \lambda)=(1.05,\,1,\,0.5)$ and for the master oscillator
		the parameters are $(1,\,1,\,0.5)$. 40000 integration steps ($0.01$) were simulated in 
		each run, and the first half was discarded from the analysis to avoid transient effects.}
\end{figure}

For the Poincar\'e system, the performance of this scheme is shown in Figure~\ref{poinc_kxyMCrec} 
with 5\% frequency mismatch between ${\cal O}_1$ and the master oscillator.
For gains slightly over 0.1 high quality PS is attained and for lower values of the gain no PS
was found. In particular for $k_x=0.02$ and $k_y=0$, PS is not achieved  ($R=0.21\pm0.09$); also for $k_x=0$ and $k_y=0.02$
the results are just as poor: $R=0.29\pm0.05$. 

\begin{figure}
		\centering
		\includegraphics[width=0.8\textwidth]{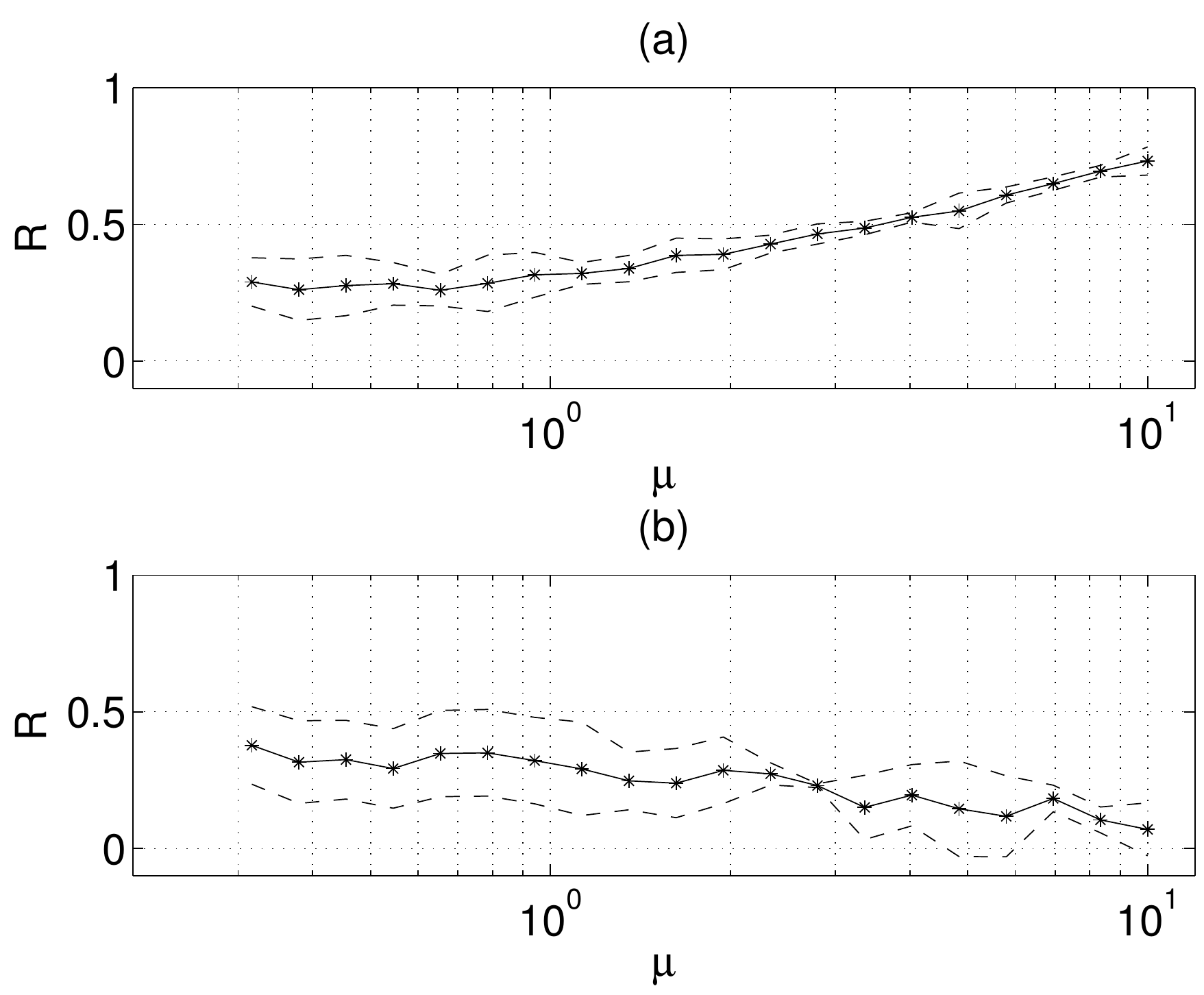}
		\caption{\label{vdp_kxyMC0p1rec}Phase coherence $R$ (\ref{pc}) for the van der Pol oscillator in Cartesian 
		coordinates with (a)~$r_x(t)=x_2(t)$, $k_x=0.1$ and $k_y=0$, (b)~$r_y(t)=y_2(t)$, $k_x=0$ and $k_y=0.1$.
		The plot shows the mean values with
		asterisks and the $\pm 2\sigma$ band with dashed lines of 10 Monte Carlo runs. The parameters for ${\cal O}_1$ are
		$(\omega,\, \mu,\, \beta)=(1.1,\,1.1\mu,\,1.1)$ and for the master oscillator
		the parameters are $(1,\,\mu,\,1)$. The simulation settings are as for Figure~\ref{poinc_kxyMCrec}.}
\end{figure}

For the van der Pol oscillator we proceed in a similar fashion, that is the control actions have the form
$m_s(t)=k_s[s_2(t)-s_1(t)],~s~\in~\{x,\,y\}$, and are added directly to the respective equation in ${\cal O}_1$,
with $k_s=0.1$. The results are summarized in Figure~\ref{vdp_kxyMC0p1rec} for 10\% mismatch in all
parameters. Synchronization is poor when the control action is added to the second equation ($s=y$), and is
only reasonable at large values of $\mu$ for $s=x$.

\begin{figure}
		\centering
		\includegraphics[width=0.8\textwidth]{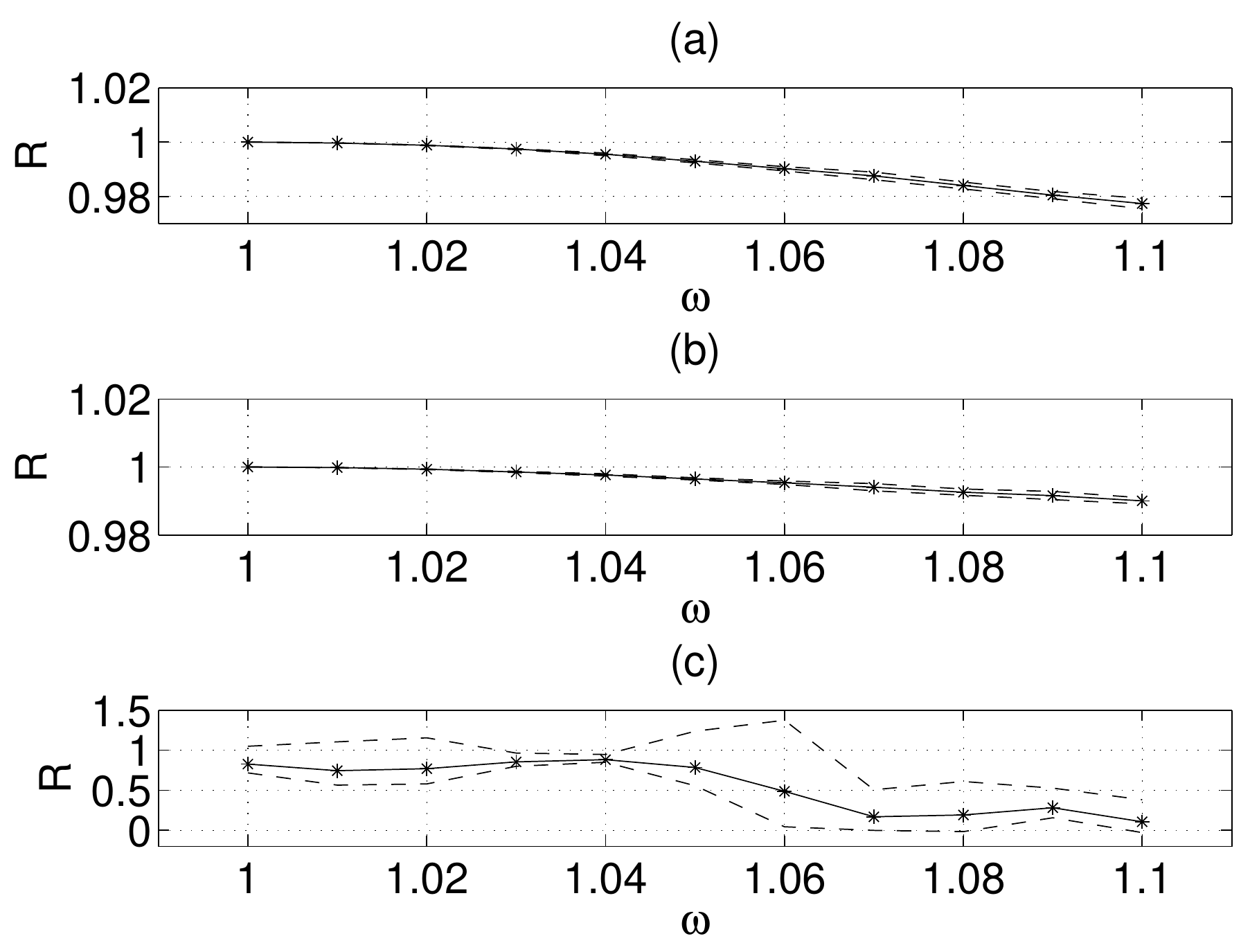}
		\caption{\label{rossl_kxyzMCrec}Phase coherence $R$ (\ref{pc}) for the R\"ossler oscillator in Cartesian 
		coordinates with $m_s=k_s[r_s(t)-s_1(t)] $ (a)~$r_x(t)=x_2(t)$ and $(k_x,\,k_y,\,k_z)=(0.4,\,0,\,0)$, 
		(b)~$r_y(t)=y_2(t)$. and $(k_x,\,k_y,\,k_z)=(0,\,0.4,\,0)$, 
		(c)~$r_z(t)=z_2(t)$. and $(k_x,\,k_y,\,k_z)=(0,\,0,\,0.4)$. Results for 10 Monte Carlo runs with random initial conditions. 
		The plots show the mean values with asterisks and the $\pm 2\sigma$ band with dashed lines. The parameters for ${\cal O}_1$ are
		$(a,\, b,\, c,\,\omega)=(0.398,\,2.0,\,4.0,\,\omega)$ and for the master oscillator
		the parameters are $(0.398,\,2.0,\,4.0,\,1.0)$. 40000 integration steps ($0.01$) were simulated in 
		each run, and the first 30000 were discarded from the analysis to avoid transient effects.} 
\end{figure}

For the R\"ossler oscillator the control action used was $m_s(t)=k_s[s_2(t)-s_1(t)],~s\in\{x,\,y,\,z\}$, with  $k_s=0.4$. 
PS was achieved quite consistently for $s=x$ and $s=y$, but more rarely for $s=z$ and only for small frequency
mismatches, as summarized in Figure~\ref{rossl_kxyzMCrec}. It was quite unexpected to find PS with $z$ as the
coupling variable, as it is known that for this variable no CS has been reported using dissipative coupling.

\begin{figure}
		\centering
		\includegraphics[width=0.8\textwidth]{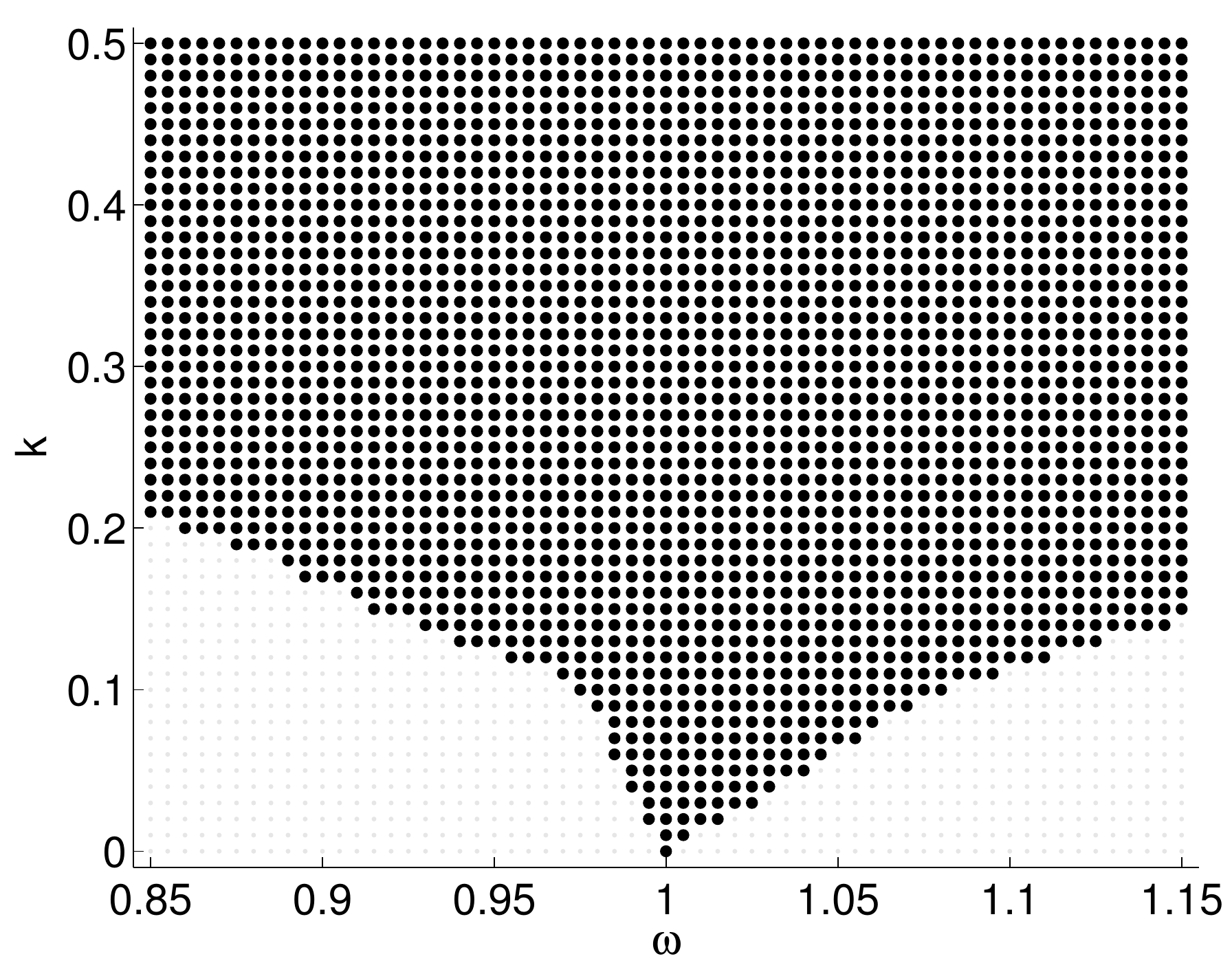} 
		\caption{\label{rossl_syncRegCart_kX1X2}Synchronization region for the R\"ossler 
system with $m_x(t)=k_x[x_2(t)-x_1(t)]$ 
		added to the first equation in (\ref{roso76}). The percentage of conditions for 
which synchronization was found (black dots) 
		is $\alpha_x=77\%$.
		The parameters for ${\cal O}_1$ are $(a,\, b,\, c,\,\omega)=(0.398,\,2.0,\,4.0,\,\omega)$ and for the master oscillator
		the parameters are $(0.398,\,2.0,\,4.0,\,1.0)$.}
\end{figure}

In addition to the phase coherence $R$, the synchronization regions on the $\omega-k_s$ plane for the R\"ossler 
oscillator were estimated using the same control actions.
Specific values of $\omega$ and $k_s$ determine a point on that plane. 
The synchronization of each point on a grid of values was verified by testing the condition $|\phi_1-\phi_2|<\pi$
directly, where $\phi_i=\tan^{-1}(y_i/x_i)$ for $i\in\{1,2\}$.
The results for the {\it best case}\, is shown in Figure~\ref{rossl_syncRegCart_kX1X2}
for a grid of pairs obtained varying $k_s\in[0,\,0.5]$ and $\omega\in [0.85,\,1.15]$ for ${\cal O}_1$ and fixed $\omega=1$
for the master oscillator. The percentage of pairs $(\omega,\,k_s)$ for which synchronization was detected
with respect of all the pairs in the grid was computed and called the $\alpha_s$ index $s~\in~\{x,\,y,\,z\}$.
For the R\"ossler system $\alpha_z \ll \alpha_y\approx\alpha_x$ (Table~\ref{Tab_obs1}). 

\subsection{Cylindrical coordinates}
\label{nrcc}

Because PS is intended here we choose $c(t)=\phi_1(t)$, $r_\phi(t)=\phi_2(t)$, 
where $\phi_2(t)$ is the phase from another similar (master) oscillator.
Connecting ${\cal S}$ to position 2 (Figure~\ref{contloop}), $e(t)=\phi_2(t)-\phi_1(t)$ and using
dissipative coupling the control action is $m_\phi(t)=k_\phi[\phi_2(t)-\phi_1(t)]$, which is a function
of {\it phase}\, error. 

For the Poincar\'e system (\ref{poip}), as before, the parameters for ${\cal O}_1$ are
$(\omega,\, p,\, \lambda)=(1.05,\,1,\,0.5)$ and for the master oscillator
the parameters are $(1,\,1,\,0.5)$. In this case high quality PS ($R=1$) is attained for a controller gain as low as $k_\phi=0.02$. 
This clearly outperforms the scheme of Section~\ref{cc}.

For the van der Pol system (\ref{vdpp}), with $k_\phi=0.5$ PS is attained with very high quality for $\mu<1$ and
even for larger values of this parameter, although with slightly decaying quality as  $\mu$ is increased.

\begin{figure}
		\centering
		\includegraphics[width=0.8\textwidth]{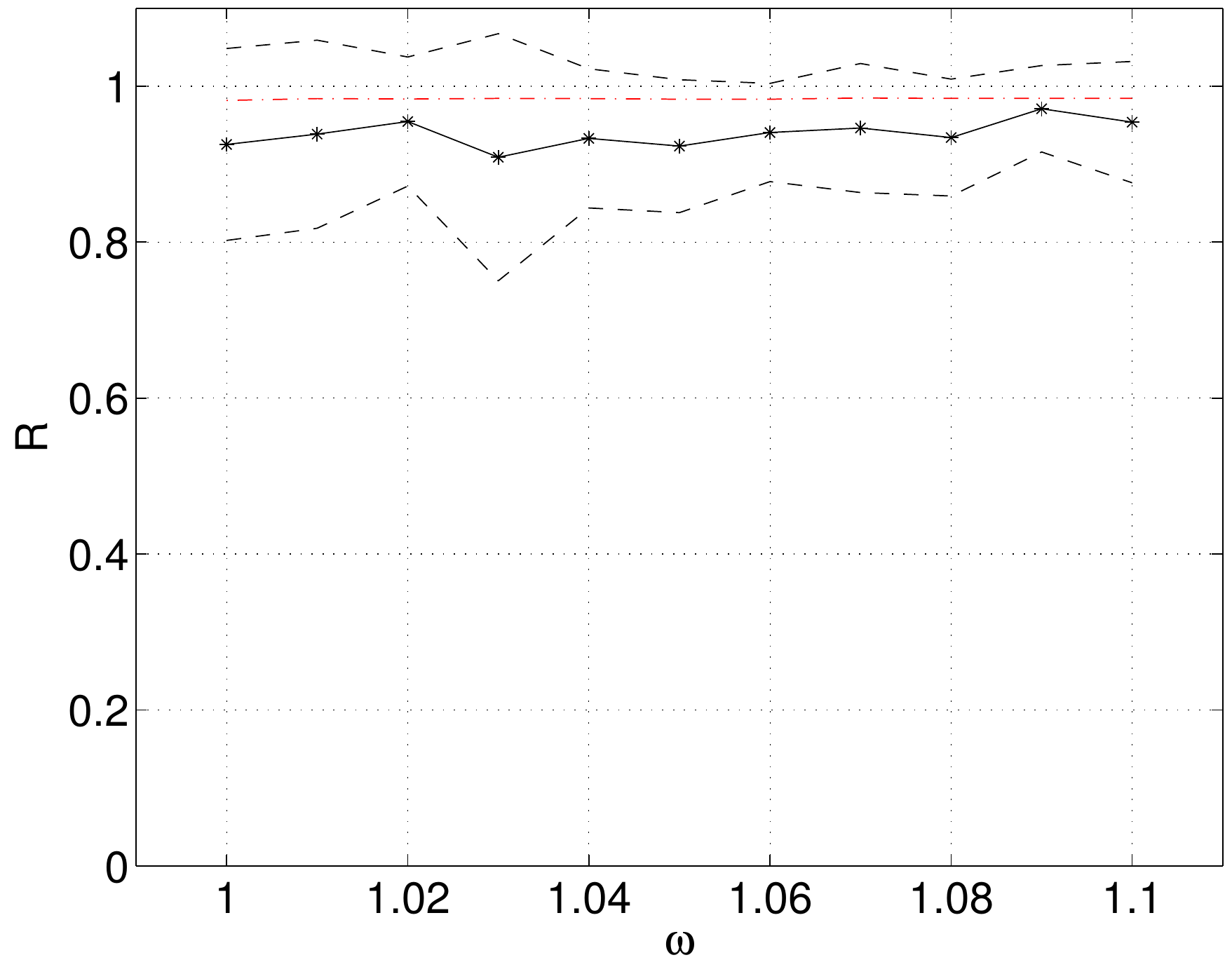}
		\caption{\label{rossl_CylMC}Phase coherence $R$ (\ref{pc}) for the R\"ossler oscillator in cylindrical 
		coordinates with $k_\phi=0.25$. Results for 10 Monte Carlo runs with random initial conditions. 
		The plots show the mean values with asterisks and the $\pm 2\sigma$ band with dashed lines. The red
		dash-dot line indicates the highest value of $R$ attained in the 10 runs. The parameters
		for the master oscillator are $(a,\, b,\, c,\,\omega)=(0.398,\,2.0,\,4.0,\,1)$ and
		for ${\cal O}_1$ are $(0.398,\,2.0,\,4.0,\,\omega)$. See caption to Figure~\ref{rossl_kxyzMCrec} for simulation settings.}
\end{figure}

For the R\"ossler system (\ref{roscyl}), $\phi_2(t)$ was generated using the same equations with
$(a,\, b,\, c,\,\omega)=(0.398,\,2.0,\,4.0,\,1)$  and the parameters for ${\cal O}_1$ are
$(a,\, b,\, c,\,\omega)=(0.398,\,2.0,\,4.0,\,\omega)$ -- notice the frequency mismatch --
and even with a low gain ($k_\phi=0.05$), $|e(t)| =|\phi_2(t)-\phi_1(t)|< \pi$ is achieved. 
For this case $R=0.92\pm0.018$ was obtained over 10 Monte Carlo runs, indicating good PS.
That value of gain is close to the threshold under which no PS is attained and is much lower than
for the case in which Cartesian coordinates are used with error in a state variable. Taking $k_\phi=0.25$
and varying the frequency $\omega$, PS is attained as shown in Figure~\ref{rossl_CylMC}. Comparing this
figure to Figure~\ref{rossl_kxyzMCrec} it is seen that although variance is much higher (probably due to 
longer transients) there is no clear deterioration with the increase of $\omega$ within the investigated
range. The maximum $R$ attained (red dash-dot line) is always very close to 0.985.

The synchronization region was also calculated for the R\"ossler system described in cylindrical coordinates,
with $m_{\phi}(t)=k_\phi[\phi_2(t)-\phi_1(t)]$ for which $\alpha_{\phi}=98\%$ (Table~\ref{Tab_obs1}).

\subsection{Control action and manipulated variable mismatch}
\label{rc}

In this section $r_\phi(t)=\phi_2(t)$ as in
Sec.\,\ref{nrcc}. The important difference here is that the main oscillator, ${\cal O}_1$, is 
simulated in Cartesian coordinates and this hampers our ability to add the control action $m_\phi(t)$
directly to the equation that determines the phase dynamics, thus resulting in CA-MV mismatch. 

In what follows the control action $m_\phi(t)=k_\phi[\phi_2(t)-\phi_1(t)]$ is
added to each dynamical equation in Cartesian coordinates, one at a time.
For the three investigated systems, the phase was computed as
$\phi_1= \tan^{-1} (y_1/x_1)$ for ${\cal O}_1$, and  analogously for $\phi_2$.

\begin{figure}
		\centering
		\includegraphics[width=0.8\textwidth]{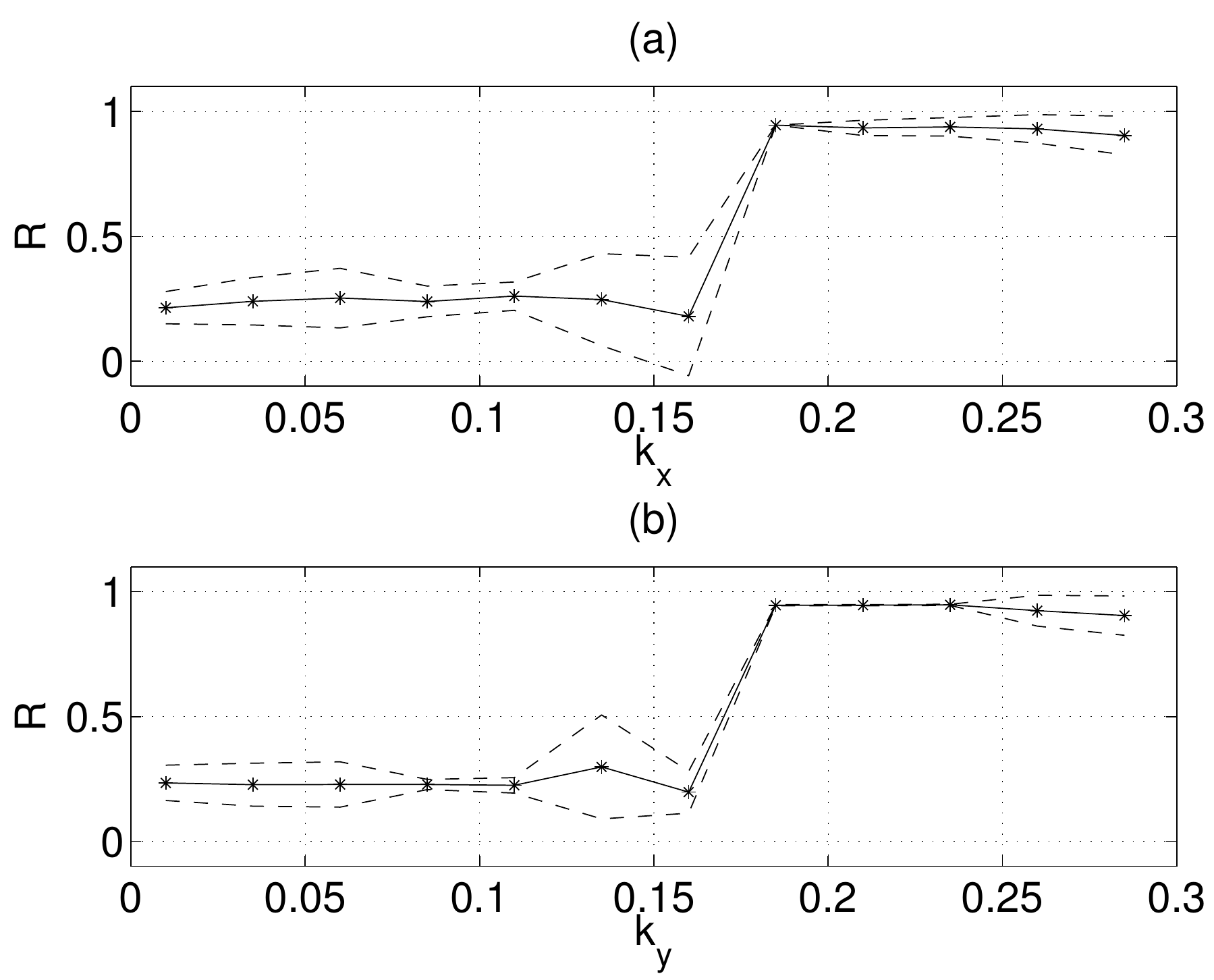}
		\caption{\label{poinc_kxyMC}Phase coherence $R$ (\ref{pc}) for the Poincar\'e oscillator in Cartesian 
		coordinates with $r_\phi(t)=\phi_2(t)$. (a)~$k_x$ and $k_y=0$, (b)~$k_x=0$ and $k_y$.
		Simulation settings  as for Figure~\ref{poinc_kxyMCrec}.}
\end{figure}

The Poincar\'e oscillator can be phase synchronized by adding the control action $m_\phi(t)$ directly 
either to the first or to the second equation with basically the same performance, as seen in 
Figure~\ref{poinc_kxyMC}. From a comparison of these results with those in Figure~\ref{poinc_kxyMCrec},
the deleterious effect of the CA-MV mismatch becomes evident.

For the van der Pol system with $m_\phi(t)=0.1[\phi_2(t)-\phi_1(t)]$, the performance is very poor in general
and PS is only attained for a limited range of values of $\mu$
and only when $m_\phi(t)$ is added to the first equation. For this system CA-MV mismatch is strongly deleterious.

\begin{figure}
		\centering
		\includegraphics[width=0.8\textwidth]{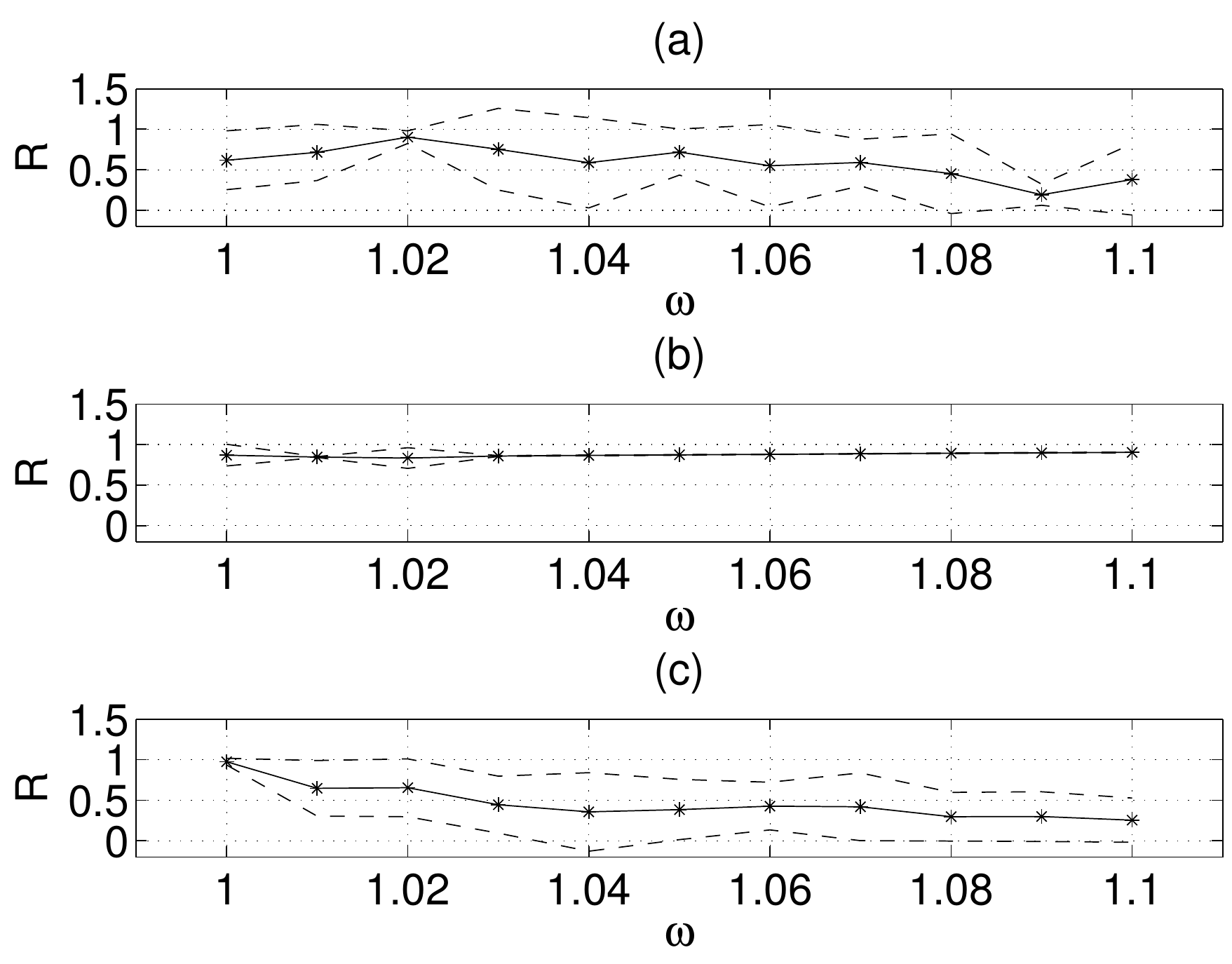}
		\caption{\label{rossl_kxyzMC}Phase coherence $R$ (\ref{pc}) for the R\"ossler oscillator in Cartesian 
		coordinates with $m_\phi(t)=k_s[\phi_2(t)-\phi_1(t)] $. (a)~$(k_x,\,k_y,\,k_z)=(0.4,\,0,\,0)$, (b)~$(k_x,\,k_y,\,k_z)=(0,\,0.4,\,0)$, 
		(c)~$(k_x,\,k_y,\,k_z)=(0,\,0,\,0.4)$. The simulation settings are as those for Figure~\ref{rossl_kxyzMCrec}. }
\end{figure}

For the R\"ossler system with $r_\phi(t)=\phi_2(t)$
rather surprisingly, PS is achieved adding the control action $m_\phi(t)=0.4[\phi_2(t)-\phi_1(t)] $
to either equation (Figure~\ref{rossl_kxyzMC}). 
For $1.0 \le \omega \le 1.02$ the {\it best}\, -- out of the 10 Monte Carlo runs -- performance of 
{\it phase coupling}\,  is similar regardless to which of the three equations $m_\phi(t)$ is added, although
adding the control action to the second equation is especially good at a higher values of  frequency mismatch.
Due to the CA-MV mismatch, the general performance is poorer than for the case shown in Figure~\ref{poinc_kxyMCrec}.

\subsection{Observability coefficients}

In order to verify if observability plays any role at all in PS, here we compute the observability
coefficients for the investigated oscillators according to \citep{let_eal/05pre}. Here we report results
for all the systems in both coordinate systems. The only situation previously considered in the literature
is that for the R\"ossler system in Cartesian coordinates, and for that system we only quote published results.

Observability matrices for the Poincar\'e oscillator are
\begin{eqnarray}
{\cal O}_\rho = 
\left[
\begin{array}{cc}
1 & 0\\
\lambda(p^2-3\rho^2) & 0 
\end{array}
\right], \hspace{0.5cm}
{\cal O}_\phi = 
\left[
\begin{array}{cc}
0 & 1\\
0 & 0 
\end{array}
\right] ,\nonumber
\end{eqnarray}

\noindent
where it is clearly seen that both matrices are singular. This confirms expectation that this system is
nonobservable if we only measure $\rho$ ($\delta_\rho=0$) or if we only measure $\phi$ ($\delta_\phi=0$), since $\rho$ cannot be found
from $\phi$ and vice versa. A very different situation happens in Cartesian coordinates, where the
observability matrices for the Poincar\'e oscillator are
\begin{eqnarray}
{\cal O}_x (\bx) & = & 
\left[
\begin{array}{cc}
1 & 0\\
\lambda(p^2-y^2-3x^2) & -\omega-2\lambda xy  \\
\end{array}
\right], \nonumber \\
{\cal O}_y (\bx) & = & 
\left[
\begin{array}{cc}
0 & 1\\
\omega-2\lambda xy   & \lambda(p^2-x^2-3y^2) 
\end{array}
\right] , \nonumber
\end{eqnarray}

\noindent
which yield the coefficients shown in Table~\ref{Tab_obs1}. 
The observability matrices for the van der Pol oscillator in cylindrical coordinates are
\begin{eqnarray}
{\cal O}_\rho (\rho,\,\phi) & = & 
\left[
\begin{array}{cc}
1 & 0\\
\sin^2\phi -3A\sin^2\phi & \rho_{22}\\
\end{array}
\right], \nonumber \\
{\cal O}_\phi (\rho,\,\phi)  & = & 
\left[
\begin{array}{cc}
0 & 1\\
-2\beta \mu \rho \cos^3\phi \sin \phi    &  \phi_{22} 
\end{array}
\right] , \nonumber
\end{eqnarray}

\noindent
where $\rho_{22}=2\rho\sin\phi \cos\phi [\beta \mu \rho (\sin^2\phi -\cos^2\phi)+1]$, 
$\phi_{22} = (\beta \mu \rho^2 \cos^2 \phi-1)(\sin^2\phi -\cos^2\phi)+ 2(\beta \mu \rho^2 \cos^2 \phi)\sin^2 \phi$. 
The observability coefficients (see Table~\ref{Tab_obs1}) 
show a clear decrease in observability from the phase for the more nonlinear situation $(\mu=10)$.  
For Cartesian coordinates the matrices are
\begin{eqnarray}
{\cal O}_x (\bx) & = & 
\left[
\begin{array}{cc}
1 & 0\\
0 & \omega \\
\end{array}
\right], \nonumber \\
{\cal O}_y (\bx) & = & 
\left[
\begin{array}{cc}
0 & \omega\\
-\omega-2\mu \beta xy   & \mu(1-\beta x^2) 
\end{array}
\right] . \nonumber
\end{eqnarray}

\noindent
It is noteworthy that ${\cal O}_x (\bx)$ for the van der Pol oscillator is constant and nonsingular. This means
that there is a global diffeomorphism connecting the original space $(x,\,y)$ and $(x,\,\dot{x})$. From an observability
point of view, this is both very good and quite rare. For the more nonlinear case $\delta_y$ is greatly reduced. 

For the R\"ossler oscillator the observability coefficients are provided in Table~\ref{Tab_obs1}. The observability
matrices for the cylindrical coordinates are cumbersome and therefore are not shown. The observability  for Cartesian coordinates can be found in
the literature. Table~\ref{Tab_obs1} also shows the percentage of synchronization conditions for the Poincar\'e, van der Pol
and R\"ossler oscillators, calculated for 61 equally-spaced values of the frequency in the range $\omega\in[0.85,\,1.15]$ 
for the slave oscillator ${\cal O}_1$ and with $\omega=1$ fixed for the master oscillator. The gain $k_s$ was varied over 51
equally-spaced values in the ranges provided in the caption of Table~\ref{Tab_obs1}. Hence for each case 3111 different
situations were tested for phase synchronization.

\begin{table}[ht]
\caption{Observability coefficients computed as in \citep{let_eal/05pre}. Index $\alpha$ indicates the percentage of
situations within the investigated ranges for which PS was verified. The investigated ranges are $\omega\in[0.85,\,1.15]$
and $k_s\in[0,\,1],~s\in\{x,\,y,\,\rho,\,\phi\}$
for the Poincar\'e and van der Pol systems, and  $k_s\in[0,\,0.5],~s\in\{x,\,y,\,z,\,\rho,\,\phi\}$ for the R\"ossler system.}\vspace{0.3cm}
\label{Tab_obs1}
	\centering
		\begin{tabular}{l|ll|ll }
		\hline
     System & \multicolumn{2}{|c|}{Cartesian} & \multicolumn{2}{|c}{Cylindrical}  \\
		\hline
    
    \multirow{2}{*}{Poincar\'e} & $\delta_x=0.340$ & $\alpha_x=75\%$ &  $\delta_\rho=0$ & $\alpha_{\rho}=8\%$ \\
		          & $\delta_y=0.338$ & $\alpha_y=75\%$ & $\delta_\phi=0$ & $\alpha_{\phi}=98\%$  \\
 \hline
    {van der Pol} &  $\delta_x=1$ & $\alpha_x=70\%$ & $\delta_\rho=0.128$ & $\alpha_{\rho}=57\%$ \\
	$\mu=1$ & $\delta_y=0.125$ & $\alpha_y=82\%$ & $\delta_\phi=0.114$ & $\alpha_{\phi}=98\%$  \\

\hline

    {van der Pol} &  $\delta_x=1$ & $\alpha_x=85\%$ & $\delta_\rho=0.036$ & $\alpha_{\rho}=14\%$ \\
	$\mu=10$ & $\delta_y=0.002$ & $\alpha_y=33\%$ & $\delta_\phi=0.001$ & $\alpha_{\phi}=77\%$ \\

\hline
    \multirow{3}{*}{R\"ossler} &  $\delta_x=0.022$ & $\alpha_x=77\%$ & $\delta_\rho=0.019$ & $\alpha_{\rho}=7\%$ \\
		          & $\delta_y=0.133$ & $\alpha_y=76\%$ & $\delta_\phi=0.001$ & $\alpha_{\phi}=98\%$  \\
		          & $\delta_z=0.006$ & $\alpha_z=4\%$ & $\delta_z=0.003$ & $\alpha_z=4\%$  \\
    \hline
\end{tabular}
\end{table}

In using observability coefficients it must be remembered that coefficients from one system should not be compared
to those of other systems, as they do not have a clear absolute interpretation. However they are useful in ranking
variables within a system and also to gain some insight. So, the observability of the dynamics from the phase is typically very
low ($\delta_\phi \ll 1$) for all systems with the exception of the van der Pol system with low nonlinearity. 
This concurs with intuition and says that it is virtually impossible to figure out
the full state of the system if only the phase is measured. Also,
typically, PS is achieved with greater ease and quality (e.g. $\alpha_{\phi}\gg\alpha_x,\,\alpha_y,\,\alpha_z,\,\alpha_{\rho}$)
when $m_\phi(t)$ is used, especially when there is no CA-MV mismatch (i.e. using $m_\phi$ in cylindrical coordinates -- see
corresponding values of $\alpha_\phi$ in Table~\ref{Tab_obs1}) or when the CA-MV mismatch is handled with the
procedure described in Sec.\,\ref{sol} (i.e. using $m_x$ and $m_y$ as defined in (\ref{choice}) -- see
Table~\ref{Tab_obs2}). So it seems fair to conclude based on these numerical results that observability
does {\it not}\, play a significant role in PS, rather the choice of controlled variable, control action and manipulated variable 
are far more fundamental.

Be it as it may, we still need to address the following question: is the relation between synchronization and observability,
at least for some systems as for instance R\"ossler, a coincidence? It is believed that the answer has two aspects.
First, as discussed before, for complete synchronization it seems reasonable that observability should play some role, so
there is no coincidence there.
Second, even in the case of PS in some cases, variables that provide good observability of the system dynamics also
are good signals for coupling. We conjecture that in such cases {\it the good coupling is a consequence of the phase information 
conveyed by the variable and is not directly related with observability proper}. In other words, such variables are probably good proxy variables
for the phase and, if so, need not provide good observability of the dynamics.

\subsection{Solving the CA-MV mismatch problem}
\label{rc2}

\begin{figure}
  \centering
  \includegraphics[width=0.8\textwidth]{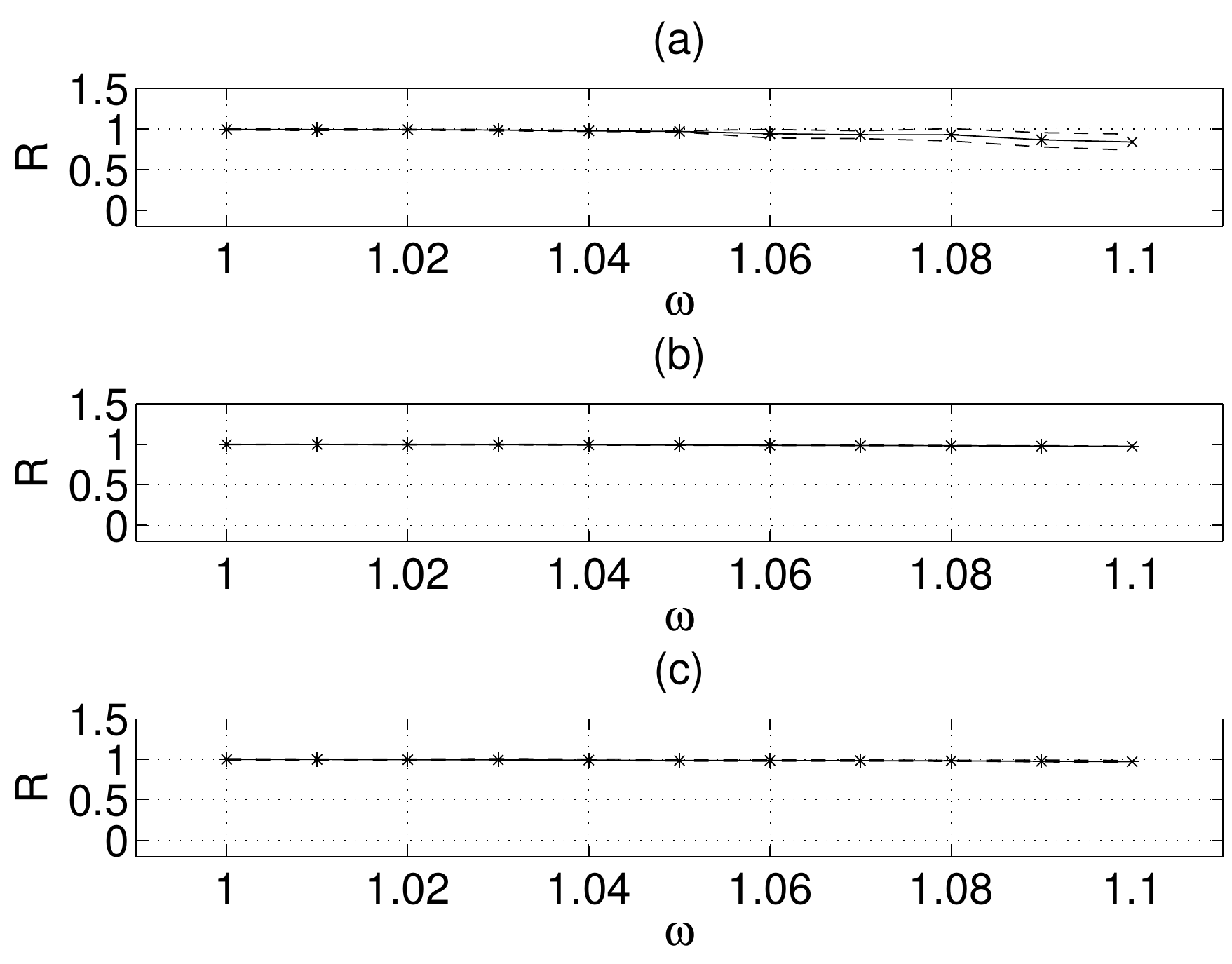}
  \caption{\label{rossl_kxkykxy_CAMVmatched}Phase coherence $R$ for the R\"ossler oscillator in Cartesian coordinates with 
(see Eq.\,\ref{choice}) (a)~$m_x=-y_1 k_\phi[r_{\phi_2}(t)-\phi_1(t)]$, $k_\phi=0.4$ and $(m_y,\,m_z)=(0,\,0)$; (b)~$m_y=x_1 k_\phi[r_{\phi_2}(t)-\phi_1(t)]$, $k_\phi=0.4$ and $(m_x,\,m_z)=(0,\,0)$; (c)~$m_x=-y_1 k_\phi[r_{\phi_2}(t)-\phi_1(t)]$, $m_y=x_1 k_\phi[r_{\phi_2}(t)-\phi_1(t)]$, $k_\phi=0.2$ and $m_z=0$. The simulation settings are as those for Figure~\ref{rossl_kxyzMCrec}.}
\end{figure}

Figure~\ref{rossl_kxkykxy_CAMVmatched} shows the result of using the control action developed
in Sec.\,\ref{sol} (see Eq.\,\ref{choice}). Notice that no control action is added to the third equation.
Comparing to previous results described for the R\"ossler system, it is clear that the procedure proposed
in Sec.\,\ref{sol} was effective in solving the CA-MV mismatch problem.

To further test the procedure developed in Sec.\,\ref{sol}, synchronization regions for the three oscillators investigated
were obtained numerically. The results are summarized in Table~\ref{Tab_obs2}.

\begin{table}[ht]
\caption{Synchronization regions for the three oscillators. Parameter values and ranges for gain and frequency are as for Table~\ref{Tab_obs1} 
but with the CA-MV mismatch problem solved using the procedure proposed
in Sec.\,\ref{sol}, hence in Cartesian coordinates. The $\alpha_x$ are computed with $m_x=-y_1 k_\phi[r_{\phi_2}(t)-\phi_1(t)]$ and $m_y=0$; 
the $\alpha_y$ are computed with $m_y=x_1 k_\phi[r_{\phi_2}(t)-\phi_1(t)]$ and $m_x=0$;  and the
$\alpha_{xy}$ were computed with $m_x=-y_1 k_\phi[r_{\phi_2}(t)-\phi_1(t)]$, $m_y=x_1 k_\phi[r_{\phi_2}(t)-\phi_1(t)]$.
In the case of the R\"ossler $m_z=0$ always.}\vspace{0.3cm}
\label{Tab_obs2}
	\centering
		\begin{tabular}{l|c}
		\hline
     System & Synchronization   \\
~& Regions \\
		\hline
   				     ~ & $\alpha_x=94\%$ \\
		Poincar\'e          &  $\alpha_y=94\%$   \\
			    ~          &  $\alpha_{xy}=97\%$   \\
		\hline
   				     ~ & $\alpha_x=94\%$ \\
		van der Pol          &  $\alpha_y=94\%$   \\
	       $\mu=1$          &  $\alpha_{xy}=97\%$   \\
		\hline
   				     ~ & $\alpha_x=92\%$ \\
		van der Pol          &  $\alpha_y=94\%$   \\
	       $\mu=10$          &  $\alpha_{xy}=97\%$   \\
		\hline
   				     ~ & $\alpha_x=65\%$ \\
		R\"ossler          &  $\alpha_y=74\%$   \\
			    ~          &  $\alpha_{xy}=92\%$   \\
    \hline
\end{tabular}
\end{table}

\begin{figure}
		\centering
	\begin{tabular}{c}
		(a) \\
		\includegraphics[width=0.8\textwidth]{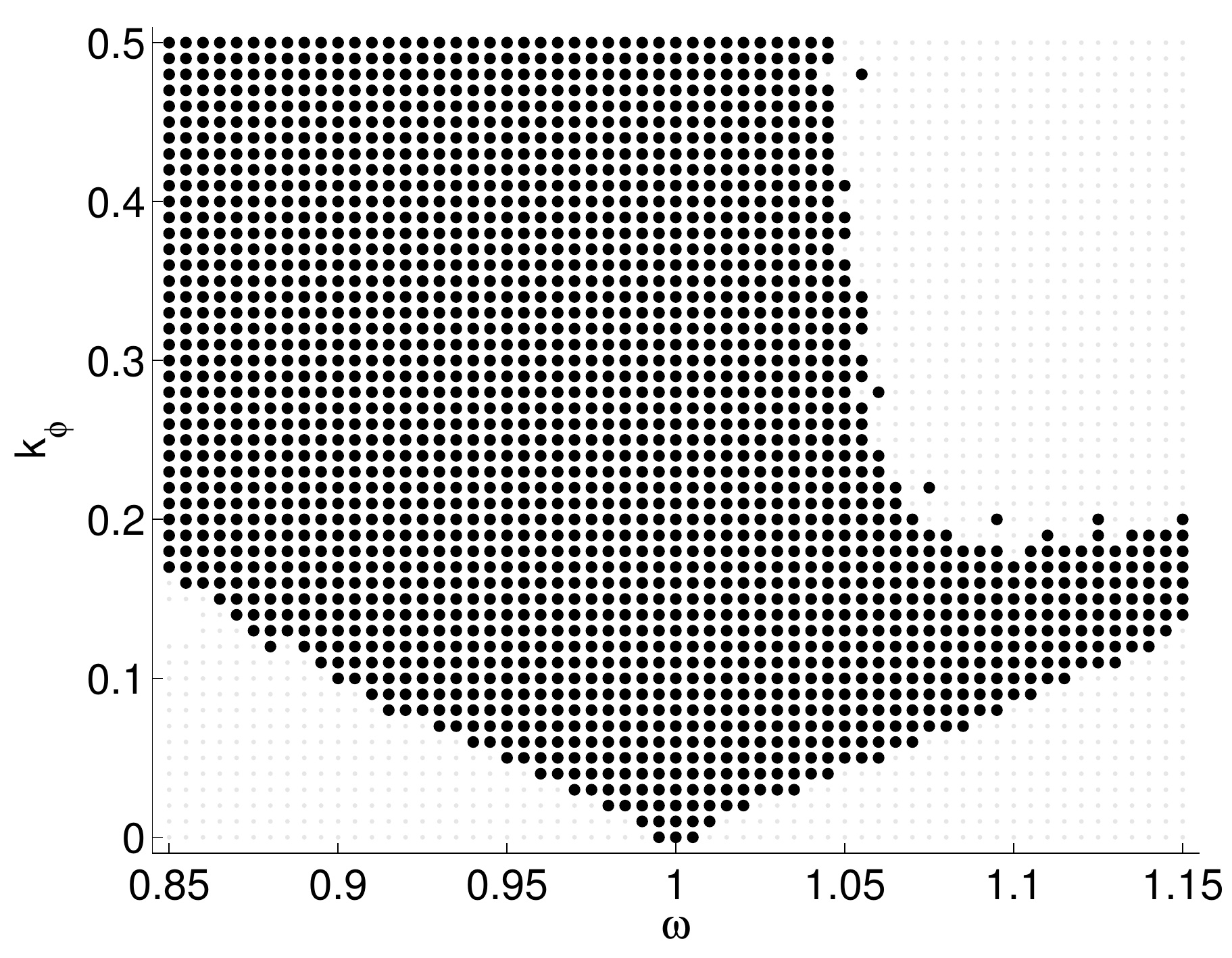} \\
		(b) \\
		\includegraphics[width=0.8\textwidth]{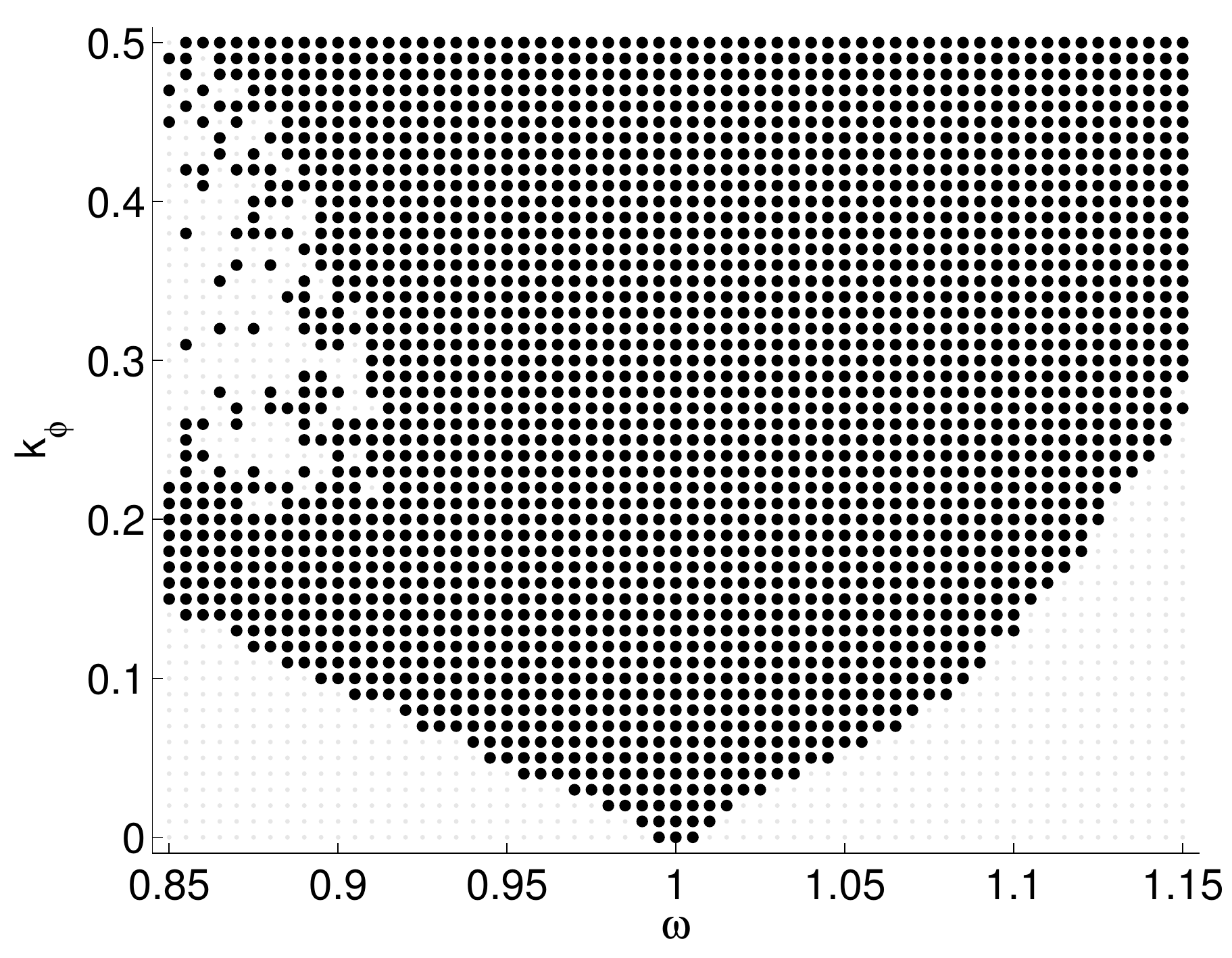} \\
	\end{tabular}
		\caption{\label{rossl_syncRegCartCAMV}Synchronization regions for the R\"ossler system with the CA-MV mismatch 
		problem solved using the procedure proposed in Sec.\,\ref{sol}
		(a)~$\alpha_x=65\%$; (b)~$\alpha_y=74\%$.
		The parameters for ${\cal O}_1$ are $(a,\, b,\, c,\,\omega)=(0.398,\,2.0,\,4.0,\,\omega)$ and for the master oscillator
		the parameters are $(0.398,\,2.0,\,4.0,\,1.0)$.}
\end{figure}

An interesting aspect revealed by such results is that $\alpha_x$ and $\alpha_y$ for the R\"ossler are
smaller than for the other oscillators. Because $m_x$ and $m_y$ defined in Sec.\,\ref{sol} are multiplied
by the state variables $-y$ and $x$, respectively, the control action has the effect of shifting the
parameter $\omega$ of the corresponding equation. In the case of the R\"ossler system in certain regions (see Figure~\ref{rossl_syncRegCartCAMV})
the slave oscillator enters the funnel regime. In such cases the lack of PS is due to the fact that a non-phase-coherent
oscillator tries to follow the phase-coherent master. Also, the assumption made in  Sec.\,\ref{sol}  that the phase
should be computable using the arctangent function is no longer valid.

\section{Discussion and Conclusions}
\label{dc}

This paper has presented a feedback control framework that is useful in analyzing master-slave
synchronization schemes. In particular, the proposed framework was used to investigate PS. It has
been argued that {\it in the case of PS}\, the key factor for successful synchronization is not the 
quality of observability provided by a state variable, but rather depends if the that variable is a good proxy 
for the phase in the case of CA-MV mismatch. When such a mismatch is overcome, for instance following
the procedure proposed in Sec.\,\ref{sol}, PS is quite simply obtained, at least for the investigated systems.
This view seems to underlie the success of phase-locked-loop schemes where the phase
detector plays a key role: receiving state variables and computing a {\it phase}\, error
rather than an error in the said variables.

For all investigated oscillators, feeding back phase had very good PS results, measured by the
coherence index and synchronization regions, but the observability
of the oscillator dynamics seen from the phase are very low in every case (see Table~\ref{Tab_obs1}).
This provides consistent counter-examples where variables with lowest observability coefficients
are best coupling variables.

An intriguing matter investigated in the paper is this: how can it be explained that for some
systems control actions based on the error of state variables instead of phase error works well?
In terms of diffusive coupling this question can be rephrased as: why, say, $m_x(t)=k_x[x_2(t)-x_1(t)]$
is sometimes very effective in attaining PS, when from a control point of view the best would be
to use  $m_\phi(t)=k_\phi[\phi_2(t)-\phi_1(t)]$? In this paper it has been conjectured that what
matters is how much of ``phase information'' is there in $x$. In other words, if $x$ is a good
proxy variable for $\phi$ then $m_x(t)$ will be a proxy for $m_\phi(t)$ and should also be effective.
There is a clear need for a way to judge the quality of proxy variables for the phase.

Another point that has been raised based on the proposed framework has been named the
{\it control action and manipulated variable (CA-MV) mismatch}. This situation happens whenever
the control action affects a manipulated variable that does not directly determine the control error.
For example, defining a phase error and computing the control action based on this is an ideal
situation from a control point of view if the aim is PS. Once the control action is available it should
affect a manipulated variable directly related to the phase dynamics, or frequency. However if it is added to
some equation in Cartesian coordinates, we face CA-MV mismatch, which is a nonideal situation.
Still, it has been shown that quality PS was attained in such a situation for the R\"ossler 
system for the $x$ and $y$ variables. We conjecture that part of this success is due to the fact
that $x$ and $y$ are good proxy variables for the phase. However this
does not explain all the details found and other aspects of the problem should be further investigated.

To illustrate this last point, consider Figure~\ref{rossl_kxyzMC} where the {\it same}\, control action
$m_\phi(t)$ was added to each equation of the R\"ossler system. Hence the coupling variable (phase) is probably not 
responsible for the differences in performance, which are probably a consequence of the dynamics associated to
each $x$, $y$ and $z$ variables. And in this respect we see that $y$ is clearly very good with $x$ and $z$
being poor. These variables perhaps suffer more intensely with the CA-MV mismatch. 

To {\it avoid}\,
the CA-MV mismatch we  represent the R\"ossler system in Cartesian coordinates and use
$m_s(t)=k_s[s_2(t)-s_1(t)],~s\in\{x,\,y,\,z\}$ (Fig.\,\ref{rossl_kxyzMCrec}). First, we must remember
that now there is no CA-MV mismatch, so there is a good chance that the performance using $x$ and
$z$ improve somewhat because these variables are sensitive to the CA-MV mismatch.
The performance in terms of $z$ has not changed significantly so any possible improvement
due to the lack of CA-MV mismatch has been negatively counter-balanced by the fact that
$z$ is a poor proxy for the phase. The performance using $y$ deteriorated slightly and that of $x$ improved greatly.
We then conjecture that $y$ is a good proxy for the phase otherwise performance would have
deteriorated for the lack of phase information. As for the $x$ variable, it must be a very good
proxy for the phase, superior to $y$. As mentioned before, this ranking of $x,\,y$ and $z$ in terms of being proxy 
variables for the phase coincides with the results in Table~\ref{Tab_obs2}.

To {\it solve}\, the CA-MV mismatch in the case of the R\"ossler system, the procedure in Sec.\,\ref{sol}
was seen to be effective. Here the phase-based control action is ``mapped'' to Cartesian coordinates.
Phase synchronization performance is then much improved. This procedure is believed to be applicable
to systems for which the phase can be estimated as the arctangent of the quotient of two state variables.

In the present investigation only systems for which the phase is well defined
were considered. This choice was made in order to be able to focus on other aspects of phase
synchronization. However, it is important to be able to extend the present results to systems
for which the phase is not readily defined. A more general definition of phase and a quantifier
of phase information in a variable are needed. These definitions are currently under investigation.

\section*{Acknowledgements}

LAA gratefully acknowledges financial support from CNPq. LF is grateful to IFMG Campus Betim for an academic leave. 
The authors thank Christophe Letellier for commenting on an early draft of this paper.

\bibliographystyle{apalike}


\begin{thebibliography}{}

\bibitem[Aguirre and Letellier, 2016]{agu_let/16}
Aguirre, L.~A. and Letellier, C. (2016).
\newblock Controllability and synchronizability: Are they related?
\newblock {\em Chaos, Solitons \& Fractals}, 83:242--251.

\bibitem[Belykh et~al., 2005]{bel_eal/05}
Belykh, V.~N., Osipov, G.~V., Kuckl{\"a}nder, N., Blasius, B., and Kurths, J.
  (2005).
\newblock Automatic control of phase synchronization in coupled complex
  oscillators.
\newblock {\em Physica D}, 200(1-2):81--104.

\bibitem[Boccaletti et~al., 2002]{boc_eal/02}
Boccaletti, S., Kurths, J., Osipov, G., Valladares, D.~L., and Zhou, C.~S.
  (2002).
\newblock The synchronization of chaotic systems.
\newblock {\em Physics Reports}, 366(1-2):1--101.

\bibitem[D\"orfler and Bullo, 2014]{dor_bul/14}
D\"orfler, F. and Bullo, F. (2014).
\newblock Synchronization in complex networks of phase oscillators: a survey.
\newblock {\em Automatica}, 50:1539--1564.

\bibitem[Freitas et~al., 2005]{fre_eal/05}
Freitas, U.~S., Macau, E. E.~N., and Grebogi, C. (2005).
\newblock Using geometric control and chaotic synchronization to estimate an
  unknown model parameter.
\newblock {\em Physical Review E}, 71(4):047203.

\bibitem[Fujisaka and Yamada, 1983]{fuj_yam/83}
Fujisaka, H. and Yamada, T. (1983).
\newblock Stability theory of synchronized motion in coupled-oscillator
  systems.
\newblock {\em Progress of Theoretical Physics}, 69(1):32--47.

\bibitem[Glass and Mackey, 1988]{gla_mac/88}
Glass, L. and Mackey, M.~C. (1988).
\newblock {\em From Clocks to Chaos, The Rhythms of Life}.
\newblock Princeton University Press, Princeton, New Jersey.

\bibitem[Hermann and Krener, 1977]{her_kre/77}
Hermann, R. and Krener, Arthur, J. (1977).
\newblock Nonlinear controllability and observability.
\newblock {\em IEEE Trans. Automat. Contr.}, 22(5):728--740.

\bibitem[Hsieh and Hung, 1996]{hsi_hun/96}
Hsieh, G.~C. and Hung, J.~C. (1996).
\newblock Phase-locked loop techniques - a survey.
\newblock {\em IEEE Transactions on Industrial Electronics}, 43(6):609--615.

\bibitem[Josi\'c and Mar, 2001]{jos_mar/01}
Josi\'c, K. and Mar, D.~J. (2001).
\newblock Phase synchronization of chaotic systems with small phase diffusion.
\newblock {\em Phys. Rev. {\rm E}}, 64(056234).

\bibitem[Kalman, 1960]{kal/60}
Kalman, R.~E. (1960).
\newblock On the general theory of control systems.
\newblock In {\em Proc. First IFAC Congress Automatic Control}, pages 481--492,
  London. Butterworths.

\bibitem[Kapitaniak, 1994]{kap/94}
Kapitaniak, T. (1994).
\newblock Synchronization of chaos using continuous control.
\newblock {\em Phys. Rev. {\rm E}}, 50(2):1642--1644.

\bibitem[Kuramoto, 1975]{kur/75}
Kuramoto, Y. (1975).
\newblock Self-entrainment of a population of coupled non-linear oscillators.
\newblock In Araki, H., editor, {\em Lecture Notes in Physics vol. 39}, pages
  420--422. Springer.

\bibitem[Letellier and Aguirre, 2010]{let_agu/10}
Letellier, C. and Aguirre, L.~A. (2010).
\newblock Interplay between synchronization, observability, and dynamics.
\newblock {\em Phys. Rev. {\rm E}}, 82(016204).

\bibitem[Letellier et~al., 2005]{let_eal/05pre}
Letellier, C., Aguirre, L.~A., and Maquet, J. (2005).
\newblock Relation between observability and differential embeddings for
  nonlinear dynamics.
\newblock {\em Physical Review {\rm E}}, 71(066213).

\bibitem[Maybhate and Amritkar, 1999]{may_amr/99}
Maybhate, A. and Amritkar, R.~E. (1999).
\newblock Use of synchronization and adaptive control in parameter estimation
  from a time series.
\newblock {\em Physical Review E}, 59:284--293.

\bibitem[Mormann et~al., 2000]{mor_eal/00}
Mormann, F., Lehnertz, K., David, P., and Elger, C.~E. (2000).
\newblock Mean phase coherence as a measure for phase synchronization and its
  application to the {EEG} of epilepsy patients.
\newblock {\em Physica D: Nonlinear Phenomena}, 144:358--369.

\bibitem[Osipov et~al., 2007]{osi_eal/07}
Osipov, G.~V., Kurths, J., and Zhou, C. (2007).
\newblock {\em Synchronization in Oscillatory Networks}.
\newblock Springer, Berlin.

\bibitem[Parlitz et~al., 1996]{par_eal/96}
Parlitz, U., Junge, L., and Kocarev, L. (1996).
\newblock Synchronization-based parameter estimation from time series.
\newblock {\em Physical Review E}, 54(6):6253--6259.

\bibitem[Parlitz et~al., 2014]{par_eal/14}
Parlitz, U., Schumann-Bischoff, J., and Luther, S. (2014).
\newblock Local observability of state variables and parameters in nonlinear
  modeling quantified by delay reconstruction.
\newblock {\em Chaos}, 24(024411).

\bibitem[Pecora and Carroll, 1990]{pec_car/90}
Pecora, L.~M. and Carroll, T.~L. (1990).
\newblock Synchronization in chaotic systems.
\newblock {\em Phys. Rev. Lett.}, 64(8):821--824.

\bibitem[Piqueira, 2011]{piq/11}
Piqueira, J. R.~C. (2011).
\newblock Network of phase-locking oscillators and a possible model for neural
  synchronization.
\newblock {\em Communications in Nonlinear Science and Numerical Simulation},
  16(9):3844--3854.

\bibitem[Rosenblum et~al., 1997]{ros_eal/97}
Rosenblum, M., Pikovsky, A.~S., and Kurths, J. (1997).
\newblock From phase to lag synchonization in coupled chaotic oscillators.
\newblock {\em Phys. Rev. Lett.}, 78(22):4193--4196.

\bibitem[Rosenblum et~al., 1996]{ros_eal/96a}
Rosenblum, M.~G., Pikovsky, A.~S., and Kurths, J. (1996).
\newblock Phase synchronization of chaotic oscillators.
\newblock {\em Phys. Rev. Lett.}, 76(11):1804--1807.

\bibitem[R\"{o}ssler, 1976]{ros/76}
R\"{o}ssler, O.~E. (1976).
\newblock An equation for continuous chaos.
\newblock {\em Phys. Lett.}, 57A(5):397--398.

\bibitem[{Sedigh-Sarvestani} et~al., 2012]{sed_eal/12}
{Sedigh-Sarvestani}, M., Schiff, S.~J., and Gluckman, B.~J. (2012).
\newblock Reconstructing mammalian sleep dynamics with data assimilation.
\newblock {\em PLoS Comput Biol}, 8(11):e1002788.

\bibitem[{Sendi\~na-Nadal} et~al., 2016]{sen_eal/16}
{Sendi\~na-Nadal}, I., Boccaletti, S., and Letellier, C. (2016).
\newblock Observability coefficients for predicting the class of
  synchronizability from the algebraic structure of the local oscillators.
\newblock {\em Phys. Rev. {\rm E}}, 94(042206).

\bibitem[Stone, 1992]{sto/92}
Stone, E.~F. (1992).
\newblock Frequency entrainment of a phase coherent attractor.
\newblock 163:367--374.

\bibitem[{van der Pol}, 1927]{vdp/27}
{van der Pol}, B. (1927).
\newblock Forced oscillations in a circuit with non-linear resistance.
\newblock {\em Philosophical Magazine}, 3(13):65--80.

\end{thebibliography}

\end{document}